\documentclass{article}
\usepackage{arxiv}

\usepackage[utf8]{inputenc} % allow utf-8 input
\usepackage[T1]{fontenc}    % use 8-bit T1 fonts
\usepackage{hyperref}       % hyperlinks
\usepackage{url}            % simple URL typesetting

\usepackage{booktabs}       % professional-quality tables
\usepackage{amsfonts}       % blackboard math symbols
\usepackage{nicefrac}       % compact symbols for 1/2, etc.
\usepackage{microtype}      % microtypography
\usepackage{lipsum}
\usepackage{graphicx}
\usepackage[multiple]{footmisc}
\usepackage{multirow}
\graphicspath{ {./images/} }
\usepackage{caption}
\usepackage{subcaption}
\usepackage{blindtext}
\usepackage[square,sort,comma,numbers]{natbib}

\title{A free web service for fast COVID-19 classification of chest X-Ray images}
\author{
	Jose David Bermudez Castro \\
	Department of Electrical Engineering (DEE)\\
	Pontifical Catholic University of Rio de Janeiro\\
	Rio de Janeiro, Brasil\\
	\texttt{bermudez@ele.puc-rio.br} \\
	\And
	Ricardo Rei \\
	Department of Electrical Engineering (DEE)\\
	Pontifical Catholic University of Rio de Janeiro\\
	Rio de Janeiro, Brasil\\
	\texttt{ricardo.rei@puc-rio.br} \\
	\And
	Jose Ruiz\\
	Department of Electrical Engineering (DEE)\\
	Pontifical Catholic University of Rio de Janeiro\\
	Rio de Janeiro, Brasil\\
	\texttt{joseruiz@puc-rio.br} \\
	\And
	Pedro Achanccaray Diaz\\
	Department of Electrical Engineering (DEE)\\
	Pontifical Catholic University of Rio de Janeiro\\
	Rio de Janeiro, Brasil\\
	\texttt{pachanccarayd@uni.pe} \\
	\And
	Smith Arauco Canchumuni\\
	Department of Electrical Engineering (DEE)\\
	Pontifical Catholic University of Rio de Janeiro\\
	Rio de Janeiro, Brasil\\
	\texttt{saraucoc@uni.pe} \\
	\And
	Cristian Muñoz Villalobos \\
	Department of Electrical Engineering (DEE)\\
	Pontifical Catholic University of Rio de Janeiro\\
	Rio de Janeiro, Brasil\\
	\texttt{crismunoz@puc-rio.br} \\
	\And
	Felipe Borges Coelho\\
	Department of Electrical Engineering (DEE)\\
	Pontifical Catholic University of Rio de Janeiro\\
	Rio de Janeiro, Brasil\\
	\texttt{borges@puc-rio.br} \\
	\And
	Leonardo Forero Mendoza\\
	Department of Electrical Engineering (DEE)\\
	Pontifical Catholic University of Rio de Janeiro\\
	Rio de Janeiro, Brasil\\
	\texttt{mendonza@ele.puc-rio.br} \\
	\And
	Marco Aurelio C. Pacheco\\
	Department of Electrical Engineering (DEE)\\
	Pontifical Catholic University of Rio de Janeiro\\
	Rio de Janeiro, Brasil\\
	\texttt{marco@ele.puc-rio.br} \\  
}

\begin{document}
	\maketitle
	\begin{abstract}
		\label{sec:abstract}
		% Felipe
		% Jose R
		% Smith
		
		% 1a frase - objetivos
		% 2a/3a frase - metodologias
		% 4a/5a frase - resultados
		% 6a frase conclusão
		
		The coronavirus outbreak became a major concern for society worldwide. Technological innovation and ingenuity are essential to fight COVID-19 pandemic and bring us one step closer to overcome it.
		
		Researchers over the world are working actively to find available alternatives in different fields, such as the Healthcare System,  pharmaceutic, health prevention, among others.With the rise of artificial intelligence (AI) in the last 10 years, IA-based applications have become the prevalent solution in different areas because of its higher capability, being now adopted to help combat against COVID-19.This work provides a fast detection system for COVID-19 using X-Ray images based on deep learning (DL) techniques. This system is available as a free web deployed service for fast patient classification, alleviating the high demand for standards method for COVID-19 diagnosis.
		It is constituted of two deep learning models, one to differentiate between X-Ray and non-X-Ray images based on Mobile-Net architecture, and another one to identify chest X-Ray images with characteristics of COVID-19 based on the DenseNet architecture.For real-time inference, it is provided a pair of dedicated GPUs, which reduce the computational time. The whole system can filter out non-chest X-Ray images, and detect whether the X-Ray presents characteristics of COVID-19, highlighting the most sensitive regions. 
		% Our system is ready to be viewed and easily used through a web page describing both the technical details and the results obtained. The web page is accessible by everyone and serves as a tool to assist in the classification of the disease by a unique image.
		% Also, our free available system counts with two dedicated GPUs to make inference in real-time. The entire system is capable of distinguishing chest X-Ray images from general images and detecting COVID-19 characteristics in the images highlighting the most sensitive regions. Our system is ready to be viewed and easily used through a web page describing both the technical details and the results obtained. The web page is accessible by everyone and serves as a tool to assist in the classification of the disease by a unique image.
		
		%Using a total of four different datasets, equivalent to more than 10000 images. The pulmonary X-Ray image filtering model achieved an accuracy of more than 99\% and the COVID-19 characteristic detection model achieved an accuracy value greater than 90\%.

	\end{abstract}

	% keywords can be removed

	\section{Introduction}
	\label{sec:introduction}
	The coronavirus disease 2019 (COVID-19) began as an outbreak in Wuhan, China and was declared a public health emergency of international concern by the World Health Organization (WHO) on January 30~\cite{who_2020_timeline}. COVID-19 has become a pandemic with more than 8 million confirmed cases and almost 500K deaths around 216 countries~\cite{who_2020_statistics}. There have been many efforts to fight COVID-19 around the world such as research a vaccine against COVID-19, massive creation of ventilators as well as personal protective equipment (PPE), development of fast diagnosis systems from chest X-Ray and/or thoracic computer tomography (CT), among others in many different fields.
	
	Artificial Intelligence (AI) can be applied in many different ways to fight COVID-19. As stated in~\cite{Vaishya:2020}, between the most important applications of AI in COVID-19 pandemic are: early detection of the infection from medical imaging, monitoring and prediction of the spread of the virus and mortality rates, tracing the virus by identifying hot spots, and development of drugs and vaccines by helping in clinical trials. In São Paulo, Brazil, regional authorities in cooperation with telephone companies are employing the Social Isolation Index (SII)~\cite{inloco_mapa_2020}, which is computed based on Geo-location data, to quantify the percentage of population that stays at home. Landing AI~\cite{landing_ai_2020} developed a monitoring tool that issues an alert when anyone is less than the desired distance from another person, promoting social distancing in the streets due to its effectiveness to slow down the COVID-19 spread. On the other hand, machine learning methods have been applied to predict the number of active cases around the world~\cite{linton_2020}, for fast COVID-19 detection from CT~\cite{Gozes:2020} and X-Ray~\cite{Cohen:2020}, and to develop systems for detection of those who are not wearing facial masks~\cite{face_mask_2020}. 
	
	Currently, one of the most employed methods for COVID-19 diagnosis is the viral nucleic acid detection using real-time polymerase chain reaction (RT-PCR). However, testing this method in thousands of suspenseful patients it is very delayed. To address this drawback, efforts have been made using AI to detect patients with COVID-19 in a very fast time through medical imaging technologies like CT and X-Ray images. It is because, COVID-19 causes acute, highly lethal pneumonia with clinical symptoms similar to those reported for SARS‐CoV and MERS‐CoV~\cite{HUANG2020497,KHAN2020252,doi:10.1002/jmv.25728}. The most common and widely available method using this type of technologies is by chest X-Ray images because it is available in ambulatory care facilities, handy with portable X-Ray systems, and enabling rapid classification (speeding up the disease screening). In contrast, CT scanners may not be available in many underdeveloped regions, disabling its use for quick diagnoses. Despite the advantages to detect COVID-19 by X-Ray, one of the main bottlenecks is the need of specialists (radiologists) to interpret the images. Taking this ``specialists'' expertise as \emph{a prior} knowledge to build an AI system can help them for quick triage, especially as a tool to triage patients for radiologists with case overloads.
	
	\paragraph{Related Works.}
	
	Several AI-based approaches have been proposed to detect patients with COVID-19 using X-Ray images. Most of them use pre-trained models and data augmentation, mainly due to the lack of data. \cite{Cohen:2020} create a free repository~\footnote{COVID-19 repository: \url{https://github.com/ieee8023/}} of X-Ray image data collections to train machine learning approaches for COVID-19 classification. In addition, \cite{Kalkreuth:2020} published a survey on public medical imaging data resources (CT, X-Ray, MRT and others), which can be used to increase the number of X-Ray samples. Reviews of known deep learning models used to create an automatic detection of COVID-19 from X-Ray images are presented by \cite{Apostolopoulos:2020}, \cite{Ilyas:2020} and \cite{Hemdan:2020}, which used VGG-19, MobileNetV2, Inception, Xception and ResNetV2 as pre-trained models. They used a small number of X-Ray samples with confirmed COVID-19 (224, 50 and 25 respectively). \cite{Karim:2020} collected more X-Ray samples to increase the training dataset and use pre-trained models followed by highlighting class-discriminating regions using gradient-guided class activation maps(Grad-CAM)~\cite{Selvaraju:2017} for better visualization.
	
	In~\cite{Abbas:2020} was introduced a model called Decompose, Transfer and Compose (DeTracC) to classify between healthy, severe acute respiratory syndrome (SARS) and COVID-19 in X-Ray images. This network was built in tree stages: (i) features extraction, (ii) class decomposition, and (iii) classification. The authors used AlexNet and ResNet18, as pre-trained models, and Principal Component Analysis (PCA), as a decomposition class. The training was performed with 80 samples of normal X-Ray, 105 samples of COVID-19, and 11 samples of SARS, and data augmentation to increase the dataset to 1764 was employed. In~\cite{Afshar:2020} was presented a new model based on capsule networks called COVIDS-CAPS to classify between non-COVID-19 (normal, Bacterial and Viral) and COVID-19 using less number of trainable parameters compared with pre-trained models. Their results show that COVIDS-CAPS is able to predict X-Ray with COVID-19. Despite the fact that the network was trained with few X-Ray samples, which poses uncertainties regarding the network suitability for new samples.
	
	Not only AI-based research was developed (several papers using deep learning to predict COVID-19 in X-Ray images), also software and services have been created. The startup ``DarwinIA'' developed the COVID-Net~\cite{Wang:2020}, which can help to identify COVID-19 patients using chest X-Ray radiography. Also, they made a free Kubernetes Bundle for inferencing COVID-19 X-Rays images~\footnote{\url{https://www.weave.works/blog/firekube-covid-ml}}. In addition, Seoul-based medical AI software company Lunit~\footnote{Lunit : \url{https://www.lunit.io/en/covid19/}} released its AI-powered software for chest X-Ray analysis with a limitation of 20 uses per day per user.
	
	\paragraph{Contributions.}
	In this context, this work presents a free and open source system available as a web service~\footnote{\url{http://www.iacontracovid.com.br/analise-de-imagens/}} for fast detection of COVID-19 in chest X-Ray images using deep learning. This system employs deep learning models to detect COVID-19 effects on lungs from chest X-Ray, which can be used for classification, triage, and monitor COVID-19 cases.
	
	The main contributions of this work can be summarized as follows: First, two deep learning models are presented, one to differentiate between X-Ray and non-X-Ray images based on Mobile-Net architecture, and another one to detect chest X-Ray images with characteristics of COVID-19 using a network bases on dense blocks and initialized from a pre-trained ImageNet. Second, we provide a free and open-source service with dedicated GPUs to make inferences in real-time. The source code is available at~\url{https://github.com/ICA-PUC/ServiceIA-COVID-19}.
	
	The remainder of the paper divides into four sections. Section~\ref{sec:methodology} describes the methodology adopted for each part of the system. Section~\ref{sec:dataset} presents the datasets employed to train the models for X-Ray images filter and the COVID-19 classifier. Section~\ref{sec:results} shows the results obtained in our experiments for each model, and the web interface of the whole system. Finally, Section~\ref{sec:conclusion} summarizes the conclusions and next steps to be followed in the development of our system.
	
	\section{Methodology}
	\label{sec:methodology}
	
\begin{figure}[!h]
		\centering
		\includegraphics[height=0.4\textwidth]{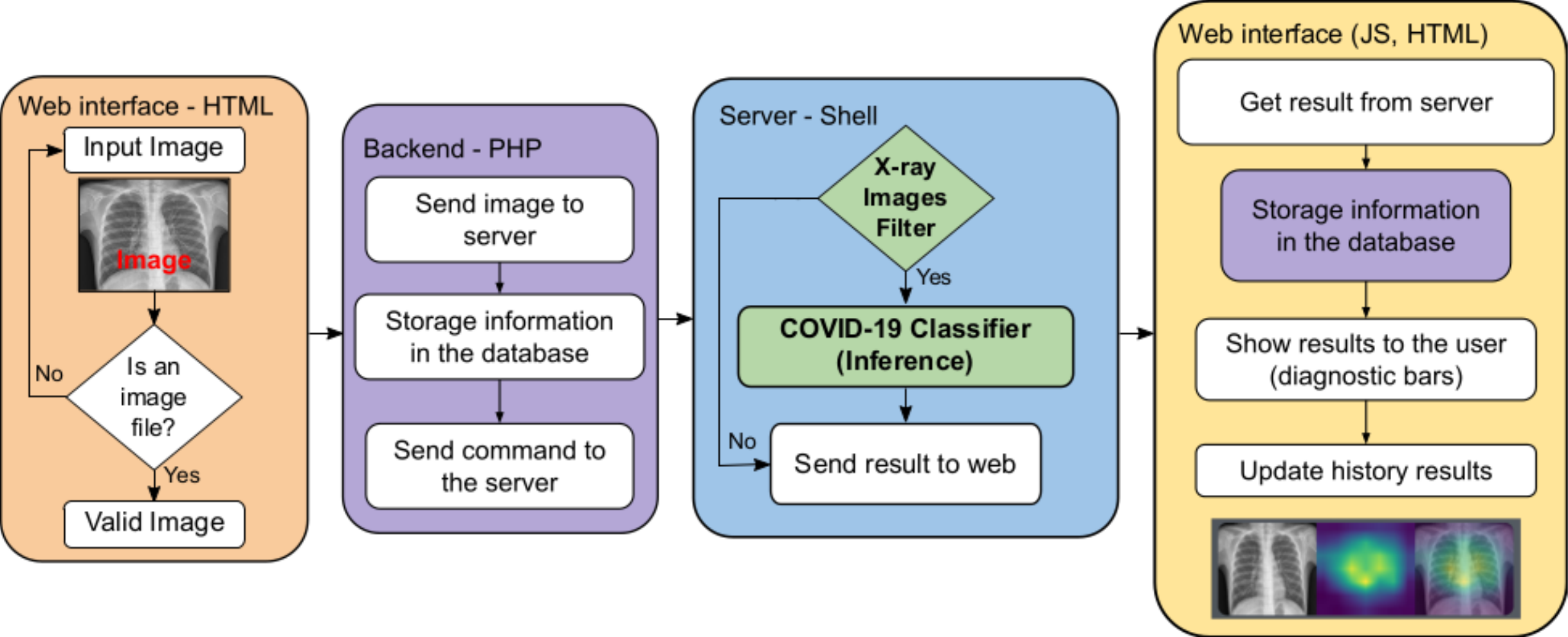}
		\caption{Inference workflow for X-ray images. Each step refers to a single abstraction layer.}
		\label{fig:diagram-web}
	\end{figure}

	This section presents the methodologies adopted for each of three parts of the project, the X-Ray image filter, the COVID-19 classifier and the Web page description. Figure~\ref{fig:diagram-web} summarizes the workflow of the system.
	%The main objective of this work is to identify pulmonary abnormality caused by COVID-19 using single X-Ray images. 
	%As mentioned before, deep learning neural networks models combined with a large image dataset of healthy and infected patients generate a neural net capable of determine whether a patient has been infected with COVID-19. Also, aiming to disseminate the research and contribute with health professionals, a system (web page) was created  to display all the research content and provide an online free service for fast classification of COVID-19. Figure~\ref{fig:diagram-web} summarizes the workflow of the system implemented in this work.
	
	The first box in Figure~\ref{fig:diagram-web}, from left to right, illustrates the beginning of the detection process. It starts with a simple extension verification to check whether the uploaded file is an image (web interface). Next, the Backend (second box) sends and stores the input image to the GPU cluster (Server), which performs the network's inferences after the Backend verifies the correct uploading of the images to it. Finally, the last box shows how the web interface stores and displays the result of the inference.
	
	% The first box on Figure~\ref{fig:diagram-web} illustrates the beginning of detection process, when the image uploaded passes through the X-Ray image verification filter (web interface). The code that processes this task runs behind the interface in the backend layer (second box on Figure 1). This layer also stores the image and sends a command to execute the inference code to the cluster. The third box (Server) presents information related to image classification with characteristics of COVID-19. Finally, the last box shows how the web interface stores and displays the result of the inference.
	
	% We divided it in three parts that together compose a single operation system. Hereafter, the methodology procedure of each part is detailed.
	
	In the following, it is described in detail the methodology employed for each constituted part of the system.
	
	\subsection{X-Ray Images Filter}
	
	Due to the input of our system to predict COVID-19 are X-Ray images from frontal AP and PA views, we first guarantee that the inputs present these characteristics to have a correct assessment. Therefore, we implement a neural network based on Mobile-Net architecture~\cite{Howard:2017} to identify whether it is a valid image (pulmonary frontal X-Ray image) or a non-valid image (rotated pulmonary X-Ray images, colored or natural images, for instance). In other words, this network operates as a filter that passes to the next process, just frontal pulmonary X-ray images from valid views.
	% The input of our system to predict COVID-19 are X-Ray images. To guarantee a correct evaluation, a neural network was implemented to identify if it is a valid image (pulmonary frontal X-Ray image) or a non-valid image (other kind of images such as rotated pulmonary X-Ray images, colored or natural images). Then, this network works as a filter that pass to the next process only frontal pulmonary X-ray images. It was implemented based on Mobile-Net architecture~\cite{Howard:2017}.
	
	%It network functions exactly as a filter that accepts only frontal pulmonary X-Ray images to the next process. This networks was built based on the MobileNet architecture.
	
	%The network was configured to receive any images as input and to generate a single value classification as output with a sigmoid activation, in which values close to one represents an acceptable image and values close to zero represent a general image.
	%The input image dimensions is resized to be $224\times224\times1$ and have as output a single neuron to which a sigmoid activation function is applied. 
	
	\subsection{COVID-19 Classifier}
\begin{figure}[!t]
		\centering
		\includegraphics[width=1.0\textwidth]{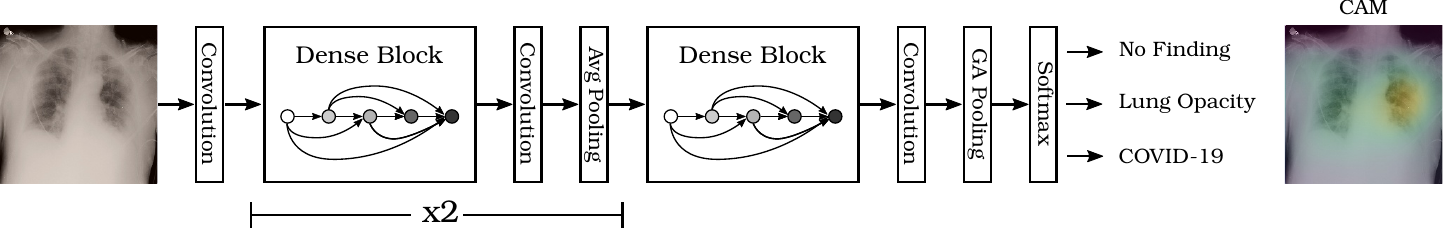}
		\caption{Description of the method followed in this work to classify X-Ray images for detection COVID-19. It is made of three Densenet Blocks, followed by a linea classifier.}
		\label{fig:covid_classifier}
	\end{figure}

	Figure~\ref{fig:covid_classifier} illustrates the processing scheme followed in this work to classify COVID-19 X-Ray images. It receives as input an X-Ray image and delivers as outputs a vector of scores indicating the probability of the image is not presenting any issues (No Finding), with opacity in Lungs (Lung Opacity), or with characteristics of COVID-19. The core of this system is a classifier built on the DenseNet~\cite{Huang:2017} network with weights initialized from a pre-trained ImageNet~\cite{deng2009imagenet} model. We selected a DenseNet-based architecture due to it has demonstrated high capability for classifying X-Ray images~\cite{rajpurkar2017chexnet, guendel2018learning, wang2018chestnet, baltruschat2019comparison, yan2019combining}. It is because the utilization of Dense blocks helps to improve the gradient flow information through the network, allowing the optimization of deeper architectures~\cite{rajpurkar2017chexnet}. 
	
	As described in Figure~\ref{fig:covid_classifier}, we use a DenseNet architecture made of three densely blocks connected in cascade, which are separated by downsampling blocks constituted of one convolution, and an average pooling layer. Finally, to the output of the last densely block, it is stacked a convolution followed by a global average pooling, and a softmax classification layer, which performs the final decision of the network.
	
	For optimizing the network parameters, we split this process into two stages due to the scarcity of COVID-19 samples. At the first stage, we train the DenseNet to discern between No Finding and Lung Opacity images, and at the second one, the model was finetuned incorporating the COVID-19 class information. During the first stage, COVID-19 samples were considered as Lung Opacity because of existing correlation between COVID-19 and Pneumonia, and processed independently in the second stage. By adopting this stratified training, we take advantage of the number of samples per class in the first stage to guarantee the learning of representative features for identifying opacities in the Lungs.  Then, in the second stage, these features are finetuned to classify whether the opacity was originated by COVID-19 or not.    
	
	\subsection{Web Page Description}
	
	The web page was created to display all the content of this research. Besides, health specialists may use the platform to support patients' classification. The web page is accessible worldwide, being available in two languages: Portuguese (original) and English.
	%Development the image analysis platform four different databases for all sub-projects were used to the X-Ray images filter and for the COVID-19 predictor.
	It also describes in detail information about the datasets, structure, technical details, and performance of classifiers. Also, there is an area on the web page where public users can experience the predicted classification using personal X-Ray images. At the end of the page, it is possible to view a 3D projection based on PCA analyses, which from a set of images samples, forms groups of images that most resemble each other.
	% Information of datasets, structure, technical details and performance of these classifiers are described in detail on the web page. Also, there is an area on the web page where public users can experience the predicted classification using personal X-Ray images. At the end of the page, it is possible to view a 3D projection based on PCA analyses, which from a set of images samples, forms groups of images that most resemble each other.
	
	\paragraph{Front-end and back-end interface - service architecture.} %In order to provide a simple solution to be utilized by any user and to quickly perform an image analysis a service available for use on the web was developed. This service can be easily updated and automatically available on the internet to be used by anyone. Below, the web site structure, the languages used with the respective libraries that facilitated its implementation are described.
	Due to the web page was developed as a free service for providing a simple solution to perform image analysis, and that can be utilized by any user, it was necessary to combine different environments programming languages to make this possible. At follow, it is detailed the web site structure, as well as the environments employed for building this service.
	
	The web page interface (called front-end) was developed in HyperText Markup Language (HTML), Cascading Style Sheets (CSS) and  JavaScript (JS), which enabled its rapid implementation by using some libraries and frameworks like: jQuery, that facilitated the implementation of JS code on the site; Dropzone\footnote{Dropzone: \url{https://www.dropzonejs.com/}}, an open source library that provides drag-and-drop file uploads with image previews; and Bootstrap 4.0\footnote{Bootstrap: \url{https://getbootstrap.com/}}, one of the world's most popular HTML, CSS and JS libraries to quickly customize responsive websites.
	
	The implemented service that maintains the operation of the website, known as backend, was developed in Hypertext Preprocessor (PHP), Python, and Shell Command Language (Sh). The Deep Learning neural network models were implemented in Keras engine, using Tensorflow as backend. The server run over a Unix operating system, which controls the running process via Bourne-again shell (Bash).
	
	% The implemented service that maintains the operation of the website, known as backend, was developed in Hypertext Preprocessor (PHP), Python and Shell Command Language (Sh). It was chosen given its simplified implementation in PHP and compatibility with all libraries and frameworks used, especially with the Dropzone library that contains large part of the implemented  code in PHP. The Deep Learning neural network models were implemented in Keras engine with backend of Tensorflow in Python and the server has a Unix operating system which required the GNU Bash or simply Bourne-again shell (Bash), which is a command language, and Unix shell whose main function is to interpret orders.
	
	%\subsubsection{Diagnostic service description}
	%In this section, the step by step of the diagnostic service is detailed by means of a flow diagram, from the insertion of the image by the user, going through all the transformations of the information, validations, and storage of the information, until the visualization of the final result of the diagnosis on the web.
	
	\section{X-Ray Dataset}
	\label{sec:dataset}
	Public data, collected from different places, was used for our implementations.
	\subsection{X-Ray Images Filter}
	For filtering frontal X-Ray images, we selected three public groups of images:
	\begin{itemize}
		\item Frontal pulmonary X-Ray~\cite{kermany2018identifying_kaggle}.
		\item Frontal non-pulmonary X-Ray~\footnote{Images were collected using Web Scraping}.
		\item Other kind of images that are not X-Ray images: PASCAL VOC~\footnote{\url{http://host.robots.ox.ac.uk/pascal/VOC/voc2012/index.html}}and Computational Vision at CALTECH dataset \footnote{\url{http://www.vision.caltech.edu/archive.html}}.
	\end{itemize}
	
	Taking into account that the performance of the classification system would be reduced if it process images inverted or rotated (90$^{\circ}$ clockwise and counterclockwise), we created a group of images with these characteristics to be identified as non-valid by the X-Ray filter. Particularly, we randomly selected a set of images from the X-Ray dataset and applied rotations of 90$^{\circ}$, 180$^{\circ}$, and 270$^{\circ}$ to each image. Table \ref{tab:dataset_xray_filter} summarizes the distribution of the dataset used for the filter network.
	
	% As the classification system should not accept inverted or rotated (90$^{\circ}$ clockwise and counterclockwise) images, a group of rotated images were created from the first group of frontal pulmonary X-Ray images. These images were obtained by applying pre-processing rotations of 90$^{\circ}$, 180$^{\circ}$ and 270$^{\circ}$ to each image. Table \ref{tab:dataset_xray_filter} shows the distribution of the dataset used for the filter network.
	
	\begin{table}[!h]
		\centering
		\begin{tabular}{llrrrr}
			\hline
			\multirow{2}{*}{Class}     & \multirow{2}{*}{Data Type}      & \multicolumn{4}{c}{\# of samples}                                                                                 \\ \cline{3-6} 
			&                                 & \multicolumn{1}{c}{Train} & \multicolumn{1}{c}{Validation} & \multicolumn{1}{c}{Test} & \multicolumn{1}{c}{Total}\\ \hline
			Valid                      & Frontal pulmonary X-Ray         & 4,400                       & 300                            & 300                     & 5,000 \\
			\multirow{3}{*}{Non-valid} & Rotated frontal pulmonary X-Ray & 4,800                       & 100                            & 100                     & 5,000 \\
			& Frontal non-pulmonary X-Ray     & 530                       & 100                            & 100                      & 730 \\
			& Not X-Ray                       & 9,800                       & 100                            & 100                     & 10,000  \\ \hline
		\end{tabular}
		\caption{Dataset distribution for the X-Ray Images Filter.}
		\label{tab:dataset_xray_filter}
	\end{table}
	
	Next, we split the dataset randomly into three sets; for training, validating, and testing the model (see Table \ref{tab:dataset_xray_filter}). Each set contains the same among of samples for each class: valid and non-valid image.  %Besides, the training samples were augmented performing random rotations of 5$^{\circ}$ and zoom of up to 10\% to increase the variety of datasets, achieving a model with better generalization.

	\subsection{COVID-19 Classifier}
\begin{figure}[!h]
		\centering
		\begin{subfigure}[b]{0.26\textwidth}
			\centering
			\includegraphics[height=0.8\textwidth]{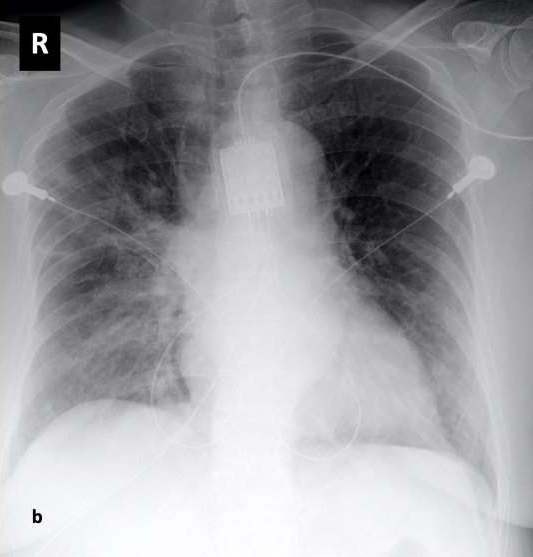}
			%\caption{}
			\label{fig:y equals x}
		\end{subfigure}
		\hfill
		\begin{subfigure}[b]{0.26\textwidth}
			\centering
			\includegraphics[height=0.8\textwidth]{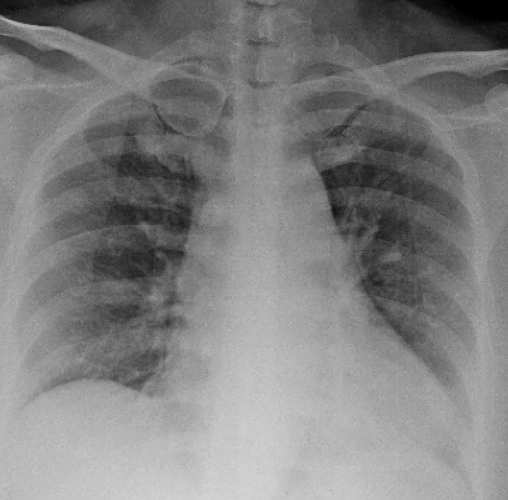}
			%\caption{}
			\label{fig:three sin x}
		\end{subfigure}
		\hfill
		\begin{subfigure}[b]{0.26\textwidth}
			\centering
			\includegraphics[height=0.8\textwidth]{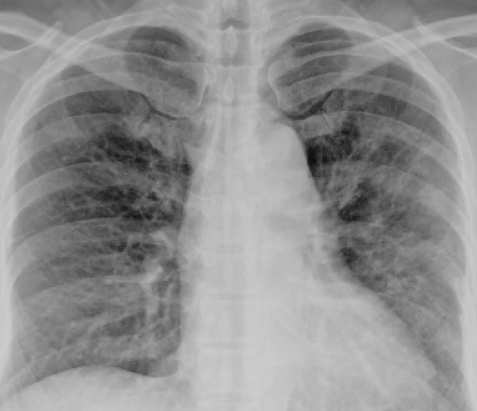}
			%\caption{}
			\label{fig:five over x}
		\end{subfigure}
		\caption{Examples of X-Ray images from patients diagnosed with COVID-19.}
		\label{fig:covid_examples}
	\end{figure}

	For detecting COVID-19 using X-Ray imagery, we selected four publicly available datasets, two associated with COVID-19 and the others with opacities in Lungs caused by different pathologies, like pneumonia, infiltration, consolidation, among others. Both COVID-19 datasets, Cohen~\cite{Cohen:2020} and \textit{Figure1-COVID}~\footnote{https://github.com/agchung/Figure1-COVID-chestxray-dataset} datasets, are community repositories updated regularly with X-Ray imagery from patients from different countries, already diagnosed or suspected of having COVID-19, and others pneumonia infections. At the date this paper was written, the Cohen dataset contained 438 X-Ray images from AP and PA views, 359 associated with COVID-19, and 79 with the others pathologies. The \textit{Figure1-COVID} dataset is a smaller one comprising 40 images, where 35 are from patients confirmed with COVID-19, three with pneumonia caused by others infections, and two with no finding pathology. There is also a set of images from patients with symptoms not confirmed as positive of COVID-19 that were filtered out. We merged the Cohen and the \textit{Figure1-COVID} dataset forming the COVID-19 dataset. In Table~\ref{tab:dataset_distribution} is summarized the distribution of samples, and in Figure~\ref{fig:covid_examples} is shown some examples of images from these datasets from patients diagnosed with COVID-19.
	
	\begin{table}[!h]
		\centering
		\begin{tabular}{lccc}
			\hline
			\multirow{2}{*}{Pathology} & \multicolumn{3}{c}{Datasets}                                                             \\ \cline{2-4} 
			& \multicolumn{1}{l}{Chest X-Ray} & \multicolumn{1}{l}{RSNA} & \multicolumn{1}{l}{COVID-19} \\ \hline
			No Finding                      & 1,150                             & 8,851                      & 4                          \\ \hline
			Lung Opacity                    & 3,100                             & 6,012                      & 82                         \\ \hline
			COVID-19                        & --                               & --                        & 394                        \\ \hline
		\end{tabular}
		\caption{Distrubtion of samples of Chest X-Ray, RSNA, and COVID-19 datasets.}
		\label{tab:dataset_distribution}
	\end{table}
	
	% \begin{figure}
	%     \centering
	%     \includegraphics[width=0.5\textwidth]{figures/classifier/covid-chestxray_hist.pdf}
	%     \caption{Distribution of samples for the COVID datasets.}
	%     \label{fig:covid_distribution}
	% \end{figure}
	
\begin{figure}[!h]
		\centering
		\begin{subfigure}[b]{0.25\textwidth}
			\centering
			\includegraphics[height=0.7\textwidth]{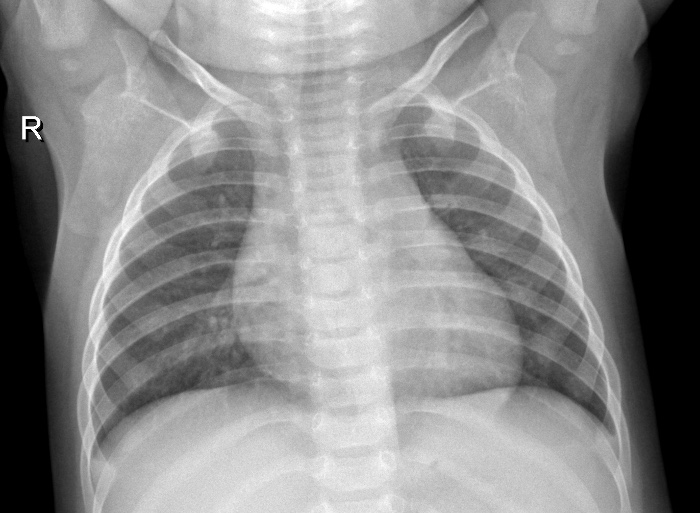}
			\caption{No Finding}
			\label{fig:ex_xray1}
		\end{subfigure}
		\hfill
		\begin{subfigure}[b]{0.25\textwidth}
			\centering
			\includegraphics[height=0.7\textwidth]{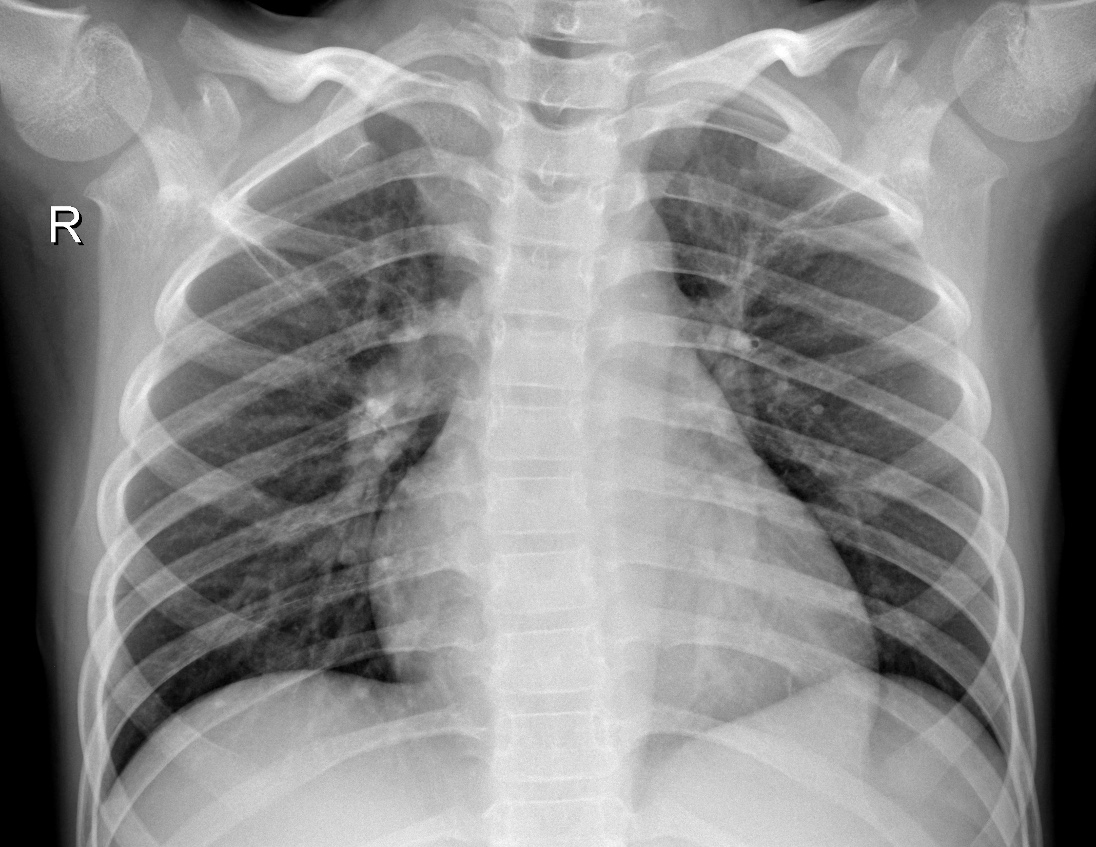}
			\caption{Virus Pneumonia}
			\label{fig:ex_xray2}
		\end{subfigure}
		\hfill
		\begin{subfigure}[b]{0.25\textwidth}
			\centering
			\includegraphics[height=0.7\textwidth]{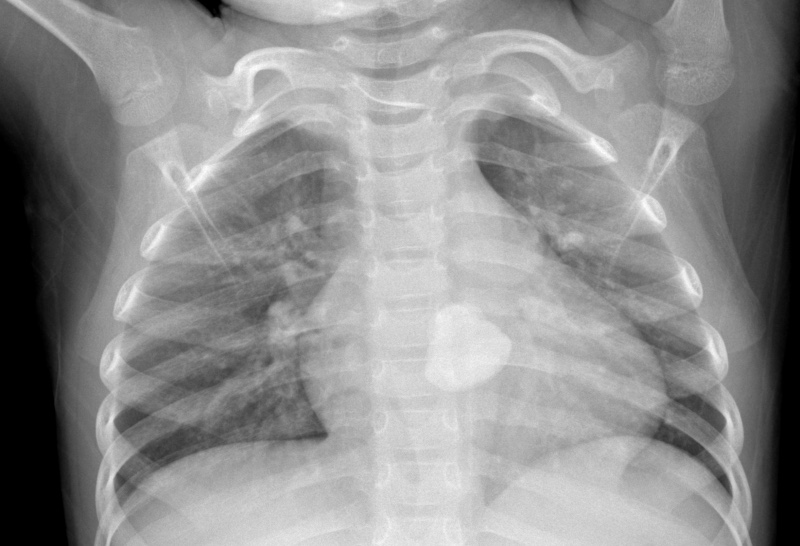}
			\caption{Bacterial Pneumonia}
			\label{fig:ex_xray3}
		\end{subfigure}
		\caption{X-Ray images from patients diagnosed as No Finding~(\subref{fig:ex_xray1}), bacterial pneumonia~(\subref{fig:ex_xray2}), and virus pneumonia~(\subref{fig:ex_xray3}).}
		\label{fig:xray_examples}
	\end{figure}

	The non-COVID-19 datasets are the Chest X-Ray Pneumonia~\cite{kermany2018labeled} and the RSNA Pneumonia detection challenge~\cite{wang2017chestx}. The Chest X-Ray Pneumonia dataset contains a total of $5,863$ PA view X-Ray images, collected from pediatric patients from one to five years old. The dataset is distributed into two categories: Normal or Pneumonia, and it is already organized into training, validation, and testing sets. Additionally, it also reported whether the pneumonia was caused by a virus or bacterial. Figure~\ref{fig:xray_examples} illustrates some examples of images registered as ``Normal'', ``Virus Pneumonia'', and ``bacterial Pneumonia''. 
	
\begin{figure}
		\centering
		\begin{subfigure}[b]{0.25\textwidth}
			\centering
			\includegraphics[height=0.8\textwidth]{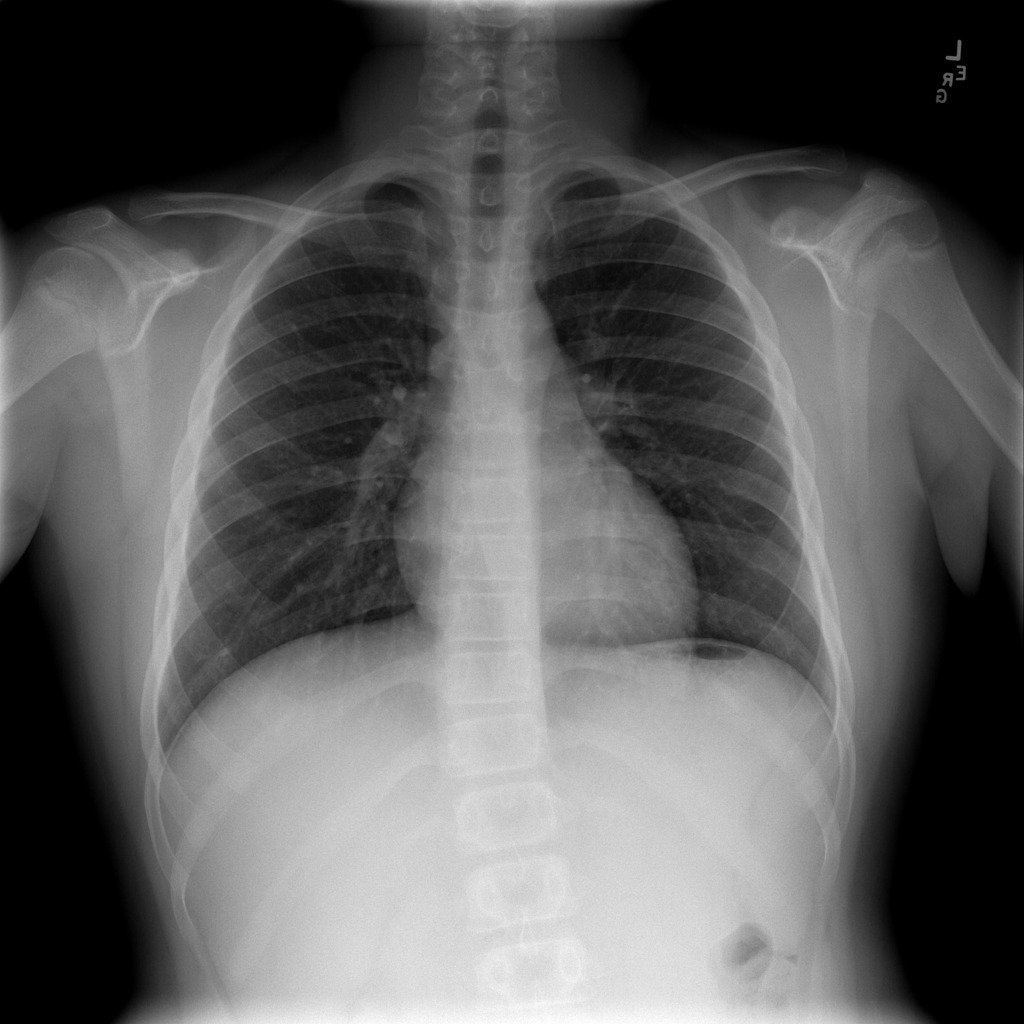}
			\caption{No Finding}
			\label{fig:exnofining1}
		\end{subfigure}
		\hfill
		\begin{subfigure}[b]{0.25\textwidth}
			\centering
			\includegraphics[height=0.8\textwidth]{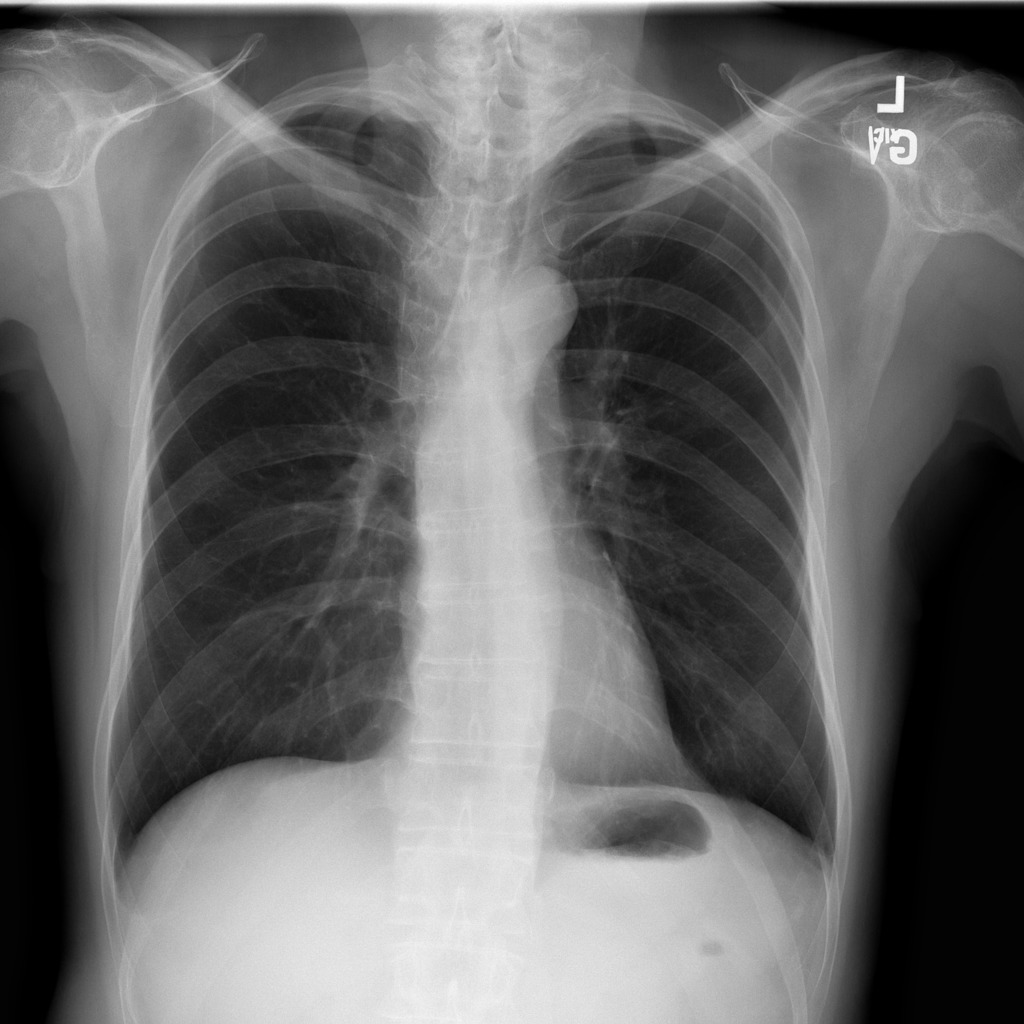}
			\caption{No Finding}
			\label{fig:exnofining2}
		\end{subfigure}
		\hfill
		\begin{subfigure}[b]{0.25\textwidth}
			\centering
			\includegraphics[height=0.8\textwidth]{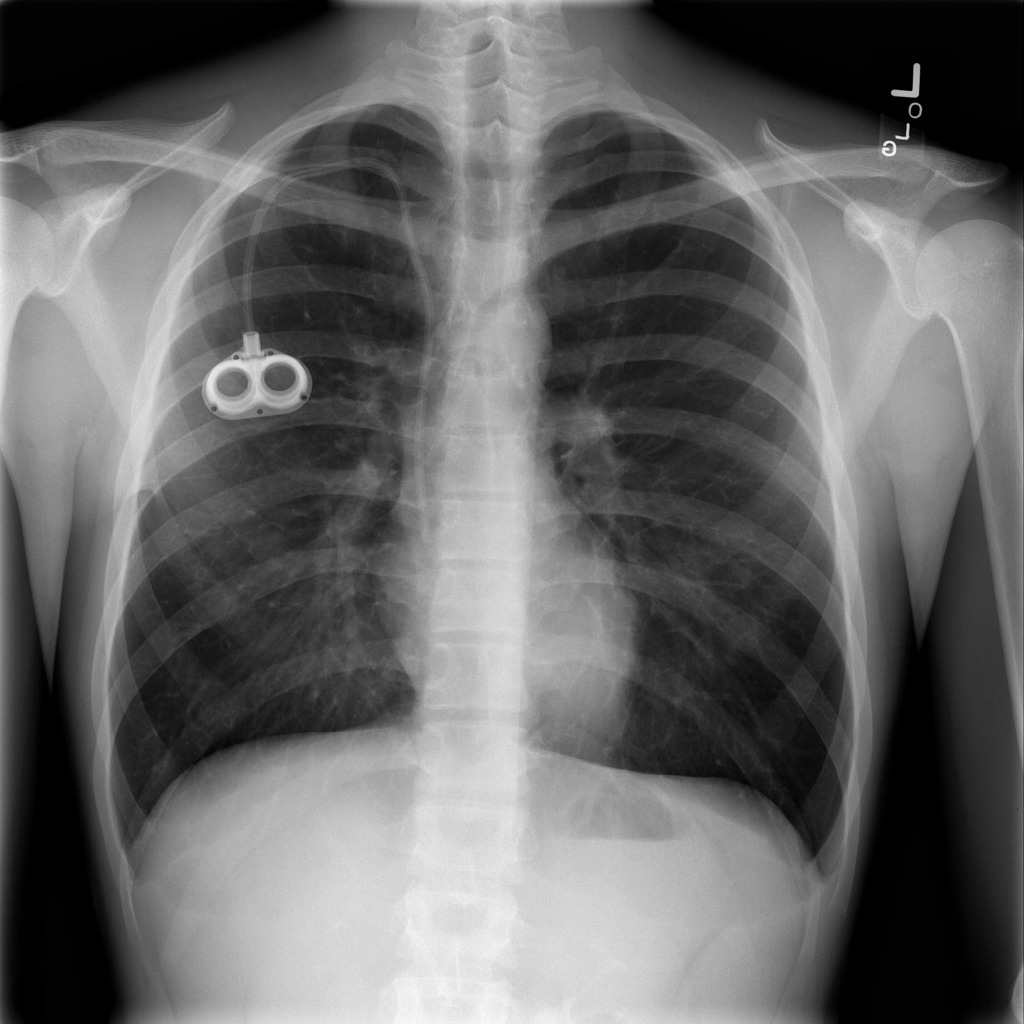}
			\caption{No Finding}
			\label{fig:exnofining3}
		\end{subfigure}
		\vfill
		
		\begin{subfigure}[b]{0.25\textwidth}
			\centering
			\includegraphics[height=0.8\textwidth]{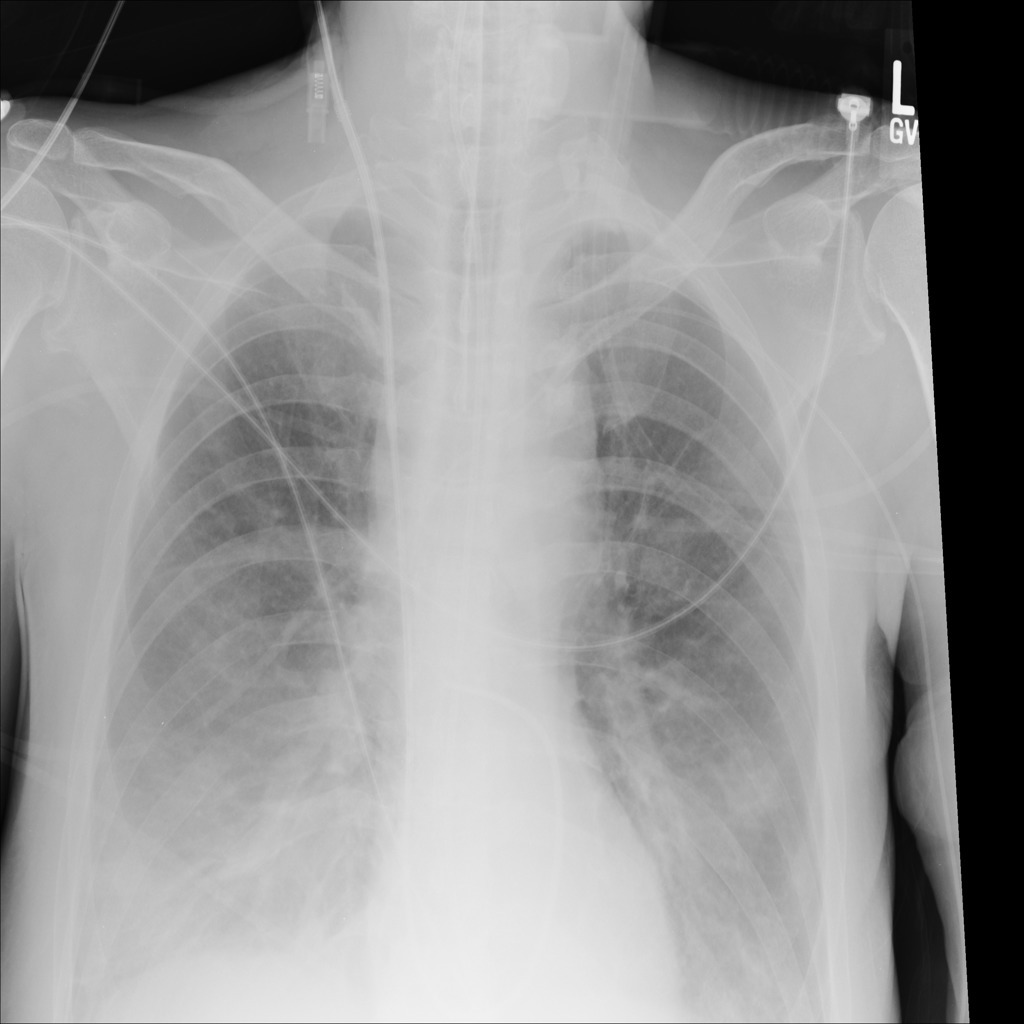}
			\caption{Lung Opacity}
			\label{fig:exopacity1}
		\end{subfigure}
		\hfill
		\begin{subfigure}[b]{0.25\textwidth}
			\centering
			\includegraphics[height=0.8\textwidth]{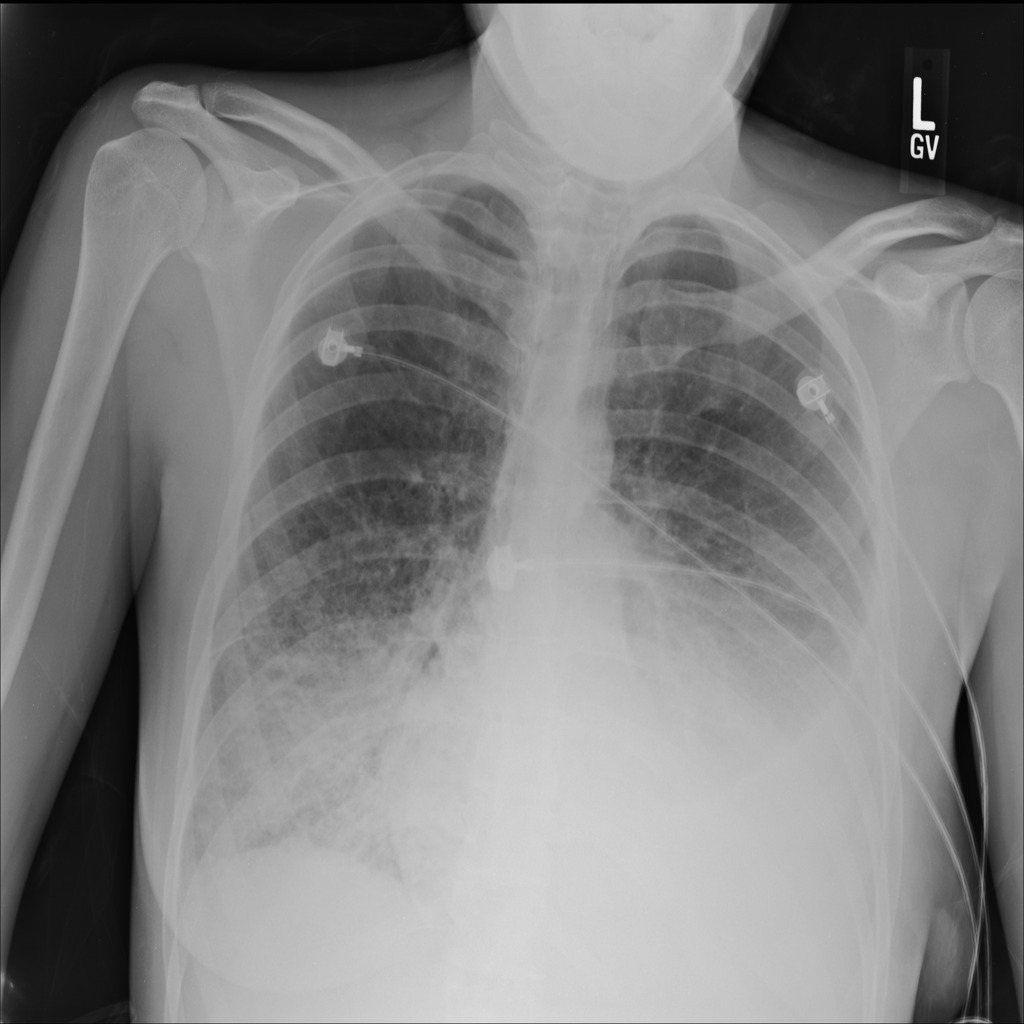}
			\caption{Lung Opacity}
			\label{fig:exopacity2}
		\end{subfigure}
		\hfill
		\begin{subfigure}[b]{0.25\textwidth}
			\centering
			\includegraphics[height=0.8\textwidth]{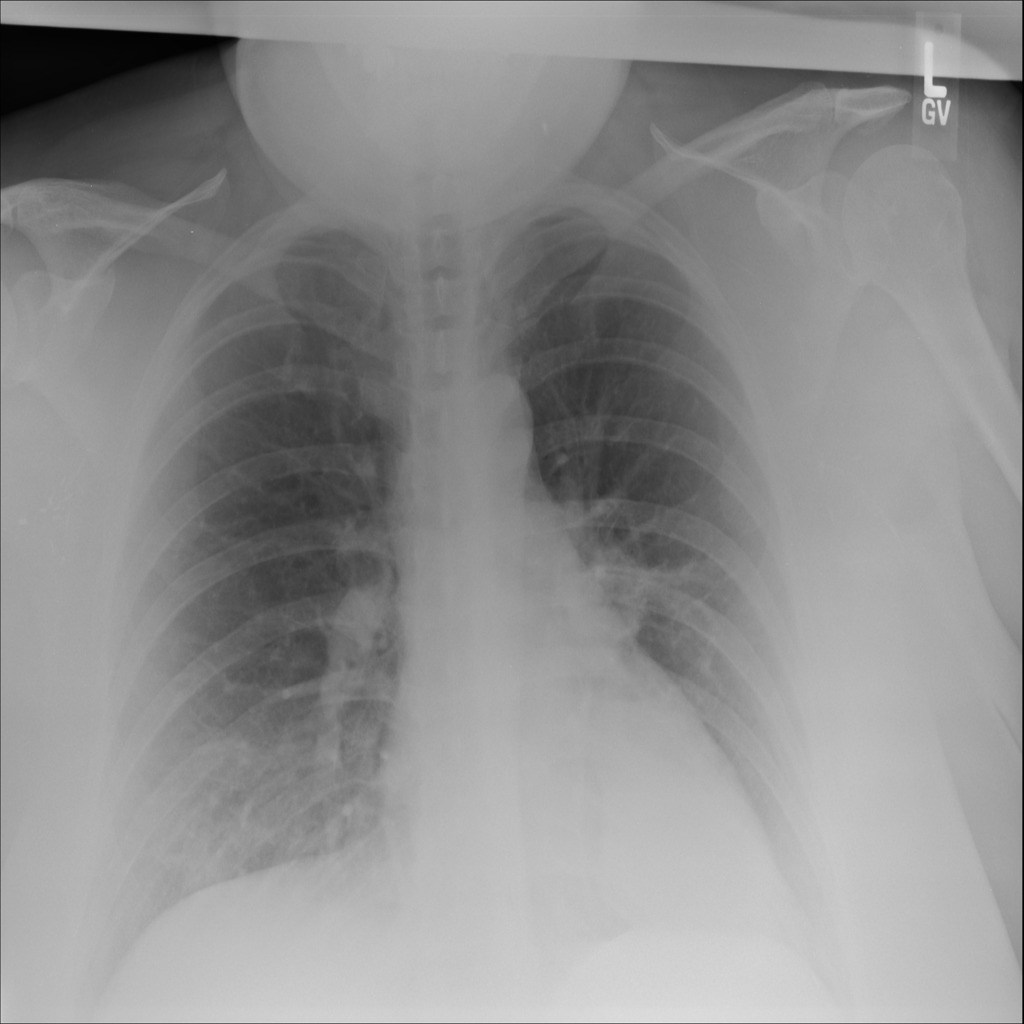}
			\caption{Lung Opacity}
			\label{fig:exopacity3}
		\end{subfigure}
		
		\caption{Examples of images from patients labeled as No Finding ~(\subref{fig:exnofining1}, \subref{fig:exnofining2}, \subref{fig:exnofining3})  and Lung Opacity~(\subref{fig:exopacity1}, \subref{fig:exopacity2}, \subref{fig:exopacity3})}
		\label{fig:rsan_exmples}
	\end{figure}

	From the RSNA dataset, we selected the images corresponding to the ``No Finding'' and ``Lung Opacity'' classes, filtering out the images rotated at (90$^{\circ}$-270$^{\circ}$) approx using the X-Ray filter. Table~\ref{tab:dataset_distribution} summarizes the final distribution of samples after performed this process. Examples of the ``No Finding'' and ``Lung Opacity'' classes are shown in Figure~\ref{fig:rsan_exmples}. 
	
	In this work, we rename the images labeled as Normal and Pneumonia to the non-Finding and Lung opacity class to have the same nomenclature of RSNA dataset, which contains images of more general cases.
	
	\newpage
	\section{Results}
	\label{sec:results}
	This section presents the experimental protocol used for training and evaluating the neural networks. Afterward, it is shown and analyzed the results obtained for each of them.
	% In this section, we present the results of the three-part of the methodology started with the details of the training the networks, and in the end the result available in the site.
	
	\subsection{X-Ray Images Filter}
	\paragraph{Experimental Protocol.}
	
	The images were resized to $224\times224\times1$ and normalized between zero to one.  Besides, the training samples were augmented by performing random rotations of  5$^{\circ}$ and zooming up to 10\%. The experiments were carried out in an NVIDIA GPU Tesla P100  of 16 GB of RAM and 3584 Cuda cores. 
	Table~\ref{tab:conf_filter_net} summarizes the metaparameters setup used for training.

	% Also, we used data augmentation over the dataset to increase a variability, into a generator class. The operation used were: (i) random rotation between $[0,5]$ and (ii) a random zoom between $[0,10]$.
	%O modelo foi treinado com as seguintes configurações:
	% The training configuration used were :
	
	\begin{table}[!h]
		\centering
		\begin{tabular}{cc}
			\hline
			Configurations & Value \\
			\hline
			Batch & 128 \\
			Steps by epoch & 5\\
			Number of epochs & 100 (maximum)\\
			Optimizer & Adam\\
			Learning Rate & 0.001 (decay a 0.5 factor each 5 epochs)\\
			Stop Criteria & Follow up the loss function (15).\\
			Metrics & Accuracy\\
			\hline
		\end{tabular}
		\caption{Training configuration used in the filter network.}
		\label{tab:conf_filter_net}
	\end{table}
	%\begin{itemize}
	%    \item batch size de 128.
	%    \item steps por epoca de 20.
	%    \item imagens totais por época de 128 $\times$ 5 de 640.
	%    \item total épocas = 100 (máximo de treino caso condições de parada não sejam alcançadas).
	%    \item optimizador do modelo = adam.
	%    \item learning rate = 0.001 (programado para baixar com um factor 0.5 a cada 5 épocas sem melhoria no val loss).
	%    \item critério de parada = 15 épocas sem melhoria no val loss.
	%    \item  pré treinada = Não.
	%    \item métricas acompanhadas = loss e acurácia (com dataset train e val)
	%\end{itemize}
	\paragraph{Result.}
	
\begin{figure}
		\begin{subfigure}[b]{0.5\textwidth}
			\centering
			\includegraphics[height=0.7\textwidth]{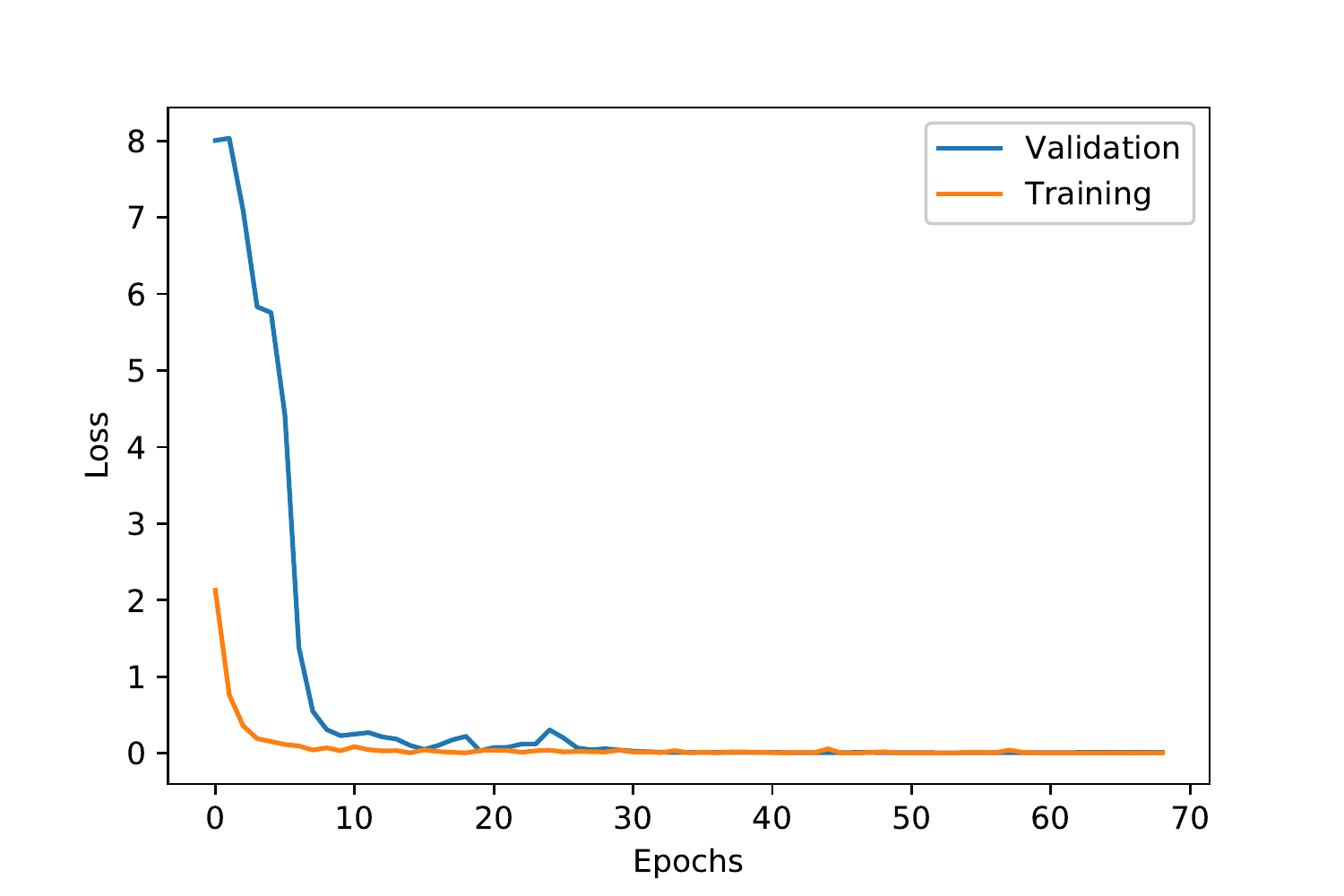}
			\caption{ }
			\label{fig:loss_linear}
		\end{subfigure}
		\begin{subfigure}[b]{0.5\textwidth}
			\centering
			\includegraphics[height=0.7\textwidth]{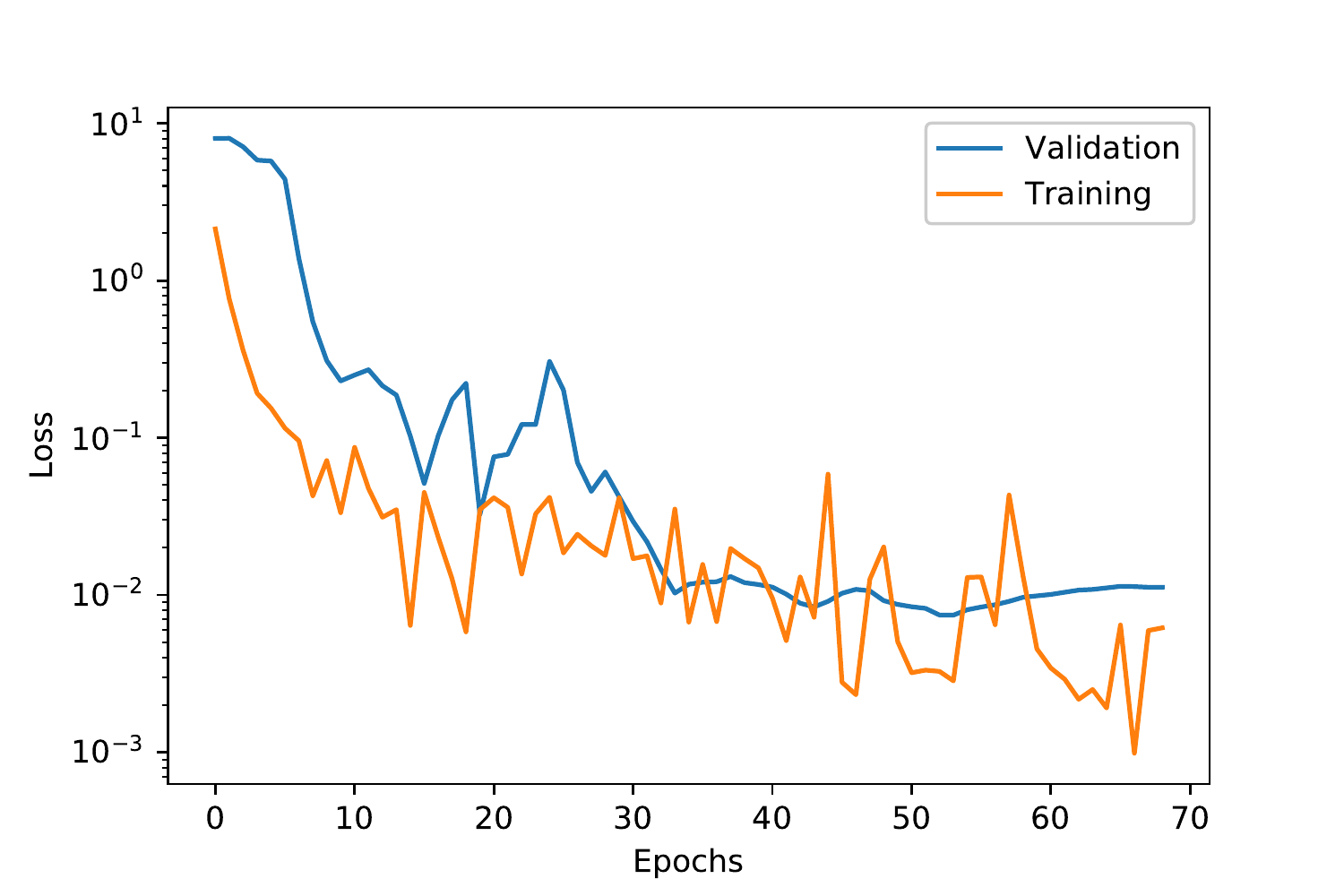}
			\caption{ }
			\label{fig:loss_logarithm}
		\end{subfigure}     
		\caption{Loss Function in (\subref{fig:loss_linear}) linear, and (\subref{fig:loss_logarithm}) logarithm scales}
		\label{fig:filter_loss}
	\end{figure}

\begin{figure}
		\centering
		\includegraphics[height=0.3\textwidth]{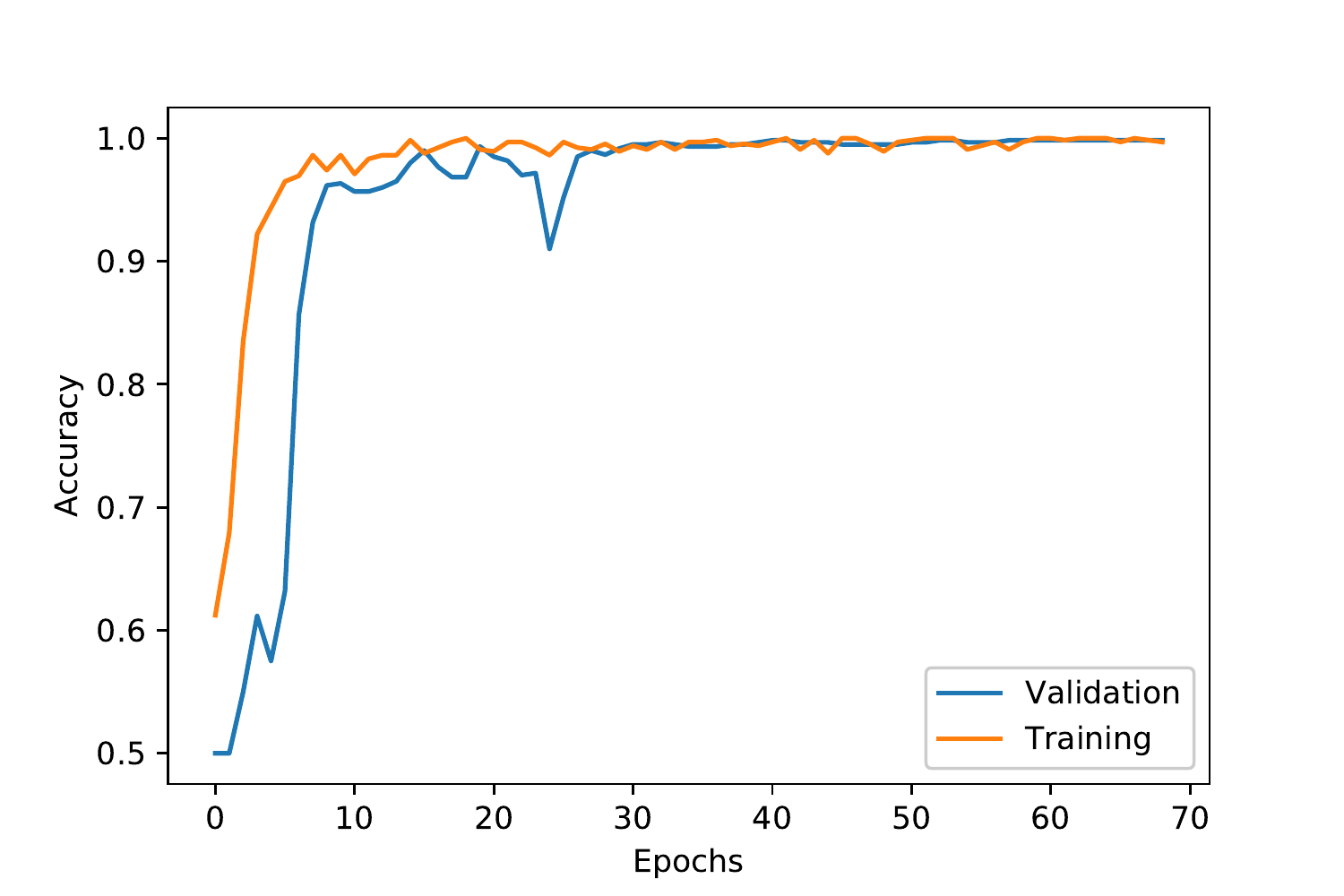}
		\caption{Training and Validation Accuracy learning curves per epoch.}
		\label{fig:filter_acc}
	\end{figure}

	The training time was 30 minutes approximately, running a total of 69 epochs, when the early-stopping was activated (see Table~\ref{tab:conf_filter_net}). Figures~\ref{fig:filter_loss} and~\ref{fig:filter_acc} shows the evolution of the loss and accuracy function over training. Notice that the network achieved the best performance at epoch 54, where the loss function was 0.0075, and the accuracy was 99.83\% in the validation set. Similar classification rates were obtained by evaluating the trained network in the testing set, reaching an accuracy of 99.3\%. 
	
	%The training was executed using GPU NVIDIA Tesla P100 with 16 GB of RAM and 3584 Cuda cores.
	% The time required for training this network was approximately 30 minutes, running a total of 69 epochs where the stop criteria was activated (see Table~\ref{tab:conf_filter_net}). Figures~\ref{fig:filter_loss} and~\ref{fig:filter_acc} shows the evolution of the loss and accuracy function over the training.
	
	%O tempo de treinamento foi de aproximadamente 30 minutos, em 54 épocas (melhor resultado de val loss) numa GPU (Nvidia Tesla P100, 16Gb RAM ). A evolução do loss e acurácia dos dados de Train e Val ao longo do treinamento foi a seguinte:
	
	% The best result was obtained in the epoch 54, with a loss function value of 0.0075 and accuracy value of 99.83\% in the validation data-set. Evaluating the trained network in the test data-set we obtained 99.3\% of accuracy. Considering that the filter network target is to deliver only frontal X-Ray images for the classifier, accuracy close to one on the test dataset is adequate.
	%which is an acceptable value for our purpose.
	%Smith
	%Na melhor época (13ª época) os resultados foram os seguintes:
	%\begin{itemize}
	%    \item dados Train - loss: 0.0028, accuracy: 1.0  
	%    \item dados Val - val loss: 0.0075 - val accuracy: 0.9983
	%\end{itemize}
	%A acurácia do modelo treinado com os dados de teste foi a seguinte:
	%\begin{itemize}
	%    \item dados Test - test accuracy: 0.993
	%\end{itemize}
	
	\subsection{COVID-19 Classifier}
	\paragraph{Experimental Protocol.}
	We first split each dataset into three groups: training, validation, and testing sets, adopting different criteria for each dataset. In the COVID-19 datasets, we selected the samples according to the patient to avoid having samples from the same patient in training and validation/testing sets, respectively. Specifically, we selected a proportion of 80\%, 10\%, and 10\% of patients for training, validation, and testing, respectively. In the RSNA dataset, the samples were split randomly following the same proportion of the COVID-19 datasets because there is no more than one image per patient. In the Chest X-Ray Pneumonia dataset was not necessary to perform this procedure as this distribution was already provided. Finally, the corresponding groups were merged to constitute the final sets. %Figure X illustrates the distributions of samples for each set after performing this procedure.
	
	All images were resized to $224\times224$ and normalized according to the mean and standard deviation of the ImageNet dataset. Due to the dataset imbalance, we both oversampled the COVID-19 images and weighted the cross-entropy loss to enhance the classification of COVID-19 imagery. We assigned the weights for each class using equation~\ref{eq:weigths},
	
	\begin{equation}
	\label{eq:weigths}
	\omega_{c_{i}} = \frac{N_{c_{i}}}{N_{c_{max}}}, 0<i<C 
	\end{equation}
	where $N_{c_{i}}$ is the number of samples for class $i$, $N_{c_{max}}$ the majority class and $C$ the number of classes.
	
	Considering the presence of images potentially labeled incorrectly, it was also applied the label smoothing as a regularization technique, setting alpha to $0.1$. It prevents the model from making predictions too confidently during training that can be reflected in poor generalization. Then, cross-entropy loss takes the form described in equation~\ref{eq:cross_entropy} 
	
	\begin{equation}
	\label{eq:cross_entropy}
	L = y_{soft} *  \log(\hat y ) * \omega_{c_{i}} + (1-y_{soft}) * \log(1-\hat y )
	\end{equation}
	where $y_{soft}=(1-\alpha)*y + \alpha / C$, and $y$ and $\hat y$ the true and predicted labels, respectively.
	
	Besides, we also augmented the training and validation samples by performing random horizontal flipping, rotation, brightness, scaling, occlusions at top of the images, cropping, and resize transformations. This procedure was carried out to avoid the network learns to maximize the inter-class differences based on the distribution of the noise that characterizes each dataset. For instance, it can be observed among the datasets, annotations like letters, arrows, captions, etc., associated with the software used for capturing the images, or/and by analysis achieved by the specialist.
	
	During training, we monitored the validation loss function applying early stopping when no improvements were observed throughout five consecutive epochs. Also, the learning rate was reduced by a factor of $0.5$ after a plateau on the validation loss. We initiated the learning rate of Adam optimizer to $1e-5$, fixing the other parameters to default values.
	
	Finally, we assessed the performance of the model quantitatively and qualitatively. The foremost was performed in terms of sensitivity and specificity metrics, which quantifies the classifier's ability to avoid false negatives and false positives, respectively. The sensitivity is defined as the ratio between the number of true positives and the sum of true positive and false negatives, and the specificity as the ratio between the true negatives and the sum of the true negatives and false positives. 
	The qualitative assessment was performed by analyzing the generated Class Activation Maps (CAM), which indicates the regions that were more relevant in the classification of the evaluated image.
	
	\paragraph{Results}
	
	Tables~\ref{tab:confusion_matrix} and \ref{tab:metrics} summarize the performance of the model in terms of the confusion matrix, and sensitivity and specificity metrics, respectively after executing the experimetens 50 times. Specifically, Table~\ref{tab:confusion_matrix} reports the sum of all classification matrix, whereas Table~\ref{tab:metrics} presents the mean and standard deviation statistics of corresponding specificity and sensitivity metrics. Results demonstrate the high capability of the model to discern between the assessed classes. In particular, it is observed that the classifier exhibits a lower rate of missclassification regarding the COVID-19 class, although the lower amount of samples. Notice the major confusion occurs between Lung Opacity and COVID-19 samples. These results are expected due to the high correlation between Lung Opacity and COVID-19 samples, where almost all patients with COVID-19 symptoms present manifestations of pneumonia.
	
	These results are reflected in the sensitivity and specificity metrics, where the classification rates are above 84.4\% in mean for all classes. However, it is necessary the collection of more COVID-19 images to improve the capability of the models for this task. For a real scenario, it is important to have a classifier with a lower rate of false positives to reduce the propagation of the virus, as well as with lower false positives to avoid classifying healthy persons.

	% CM
	% [1051   64    3]
	% [  67  901   15]
	% [   0    6   53]

	\begin{table}[!h]
		\centering
		\begin{tabular}{clccc}
			\cline{3-5}
			&                                   & \multicolumn{3}{c}{Predicted}                                                                    \\ \cline{3-5} 
			\multicolumn{1}{l}{}                         &                                   & \multicolumn{1}{l}{No Finding} & \multicolumn{1}{l}{Lung Opacity} & \multicolumn{1}{l}{COVID-19} \\ \hline
			\multicolumn{1}{c|}{\multirow{3}{*}{Actual}} & \multicolumn{1}{l|}{No Finding}   & $53821/92.5\%$    & $3552/6.1\%$     & $763/1.3\%$                            \\
			\multicolumn{1}{c|}{}                        & \multicolumn{1}{l|}{Lung Opacity}
			& $3116/6.1\%$              & $46620/91.2\%$                              & $1380/2.7\%$                            \\
			\multicolumn{1}{c|}{}                        & \multicolumn{1}{l|}{COVID-19}
			& 0             & $356/11.6\%$                                & $2712/88.4\%$                           \\ \hline
		\end{tabular}
		\caption{Sum of all Confusion Matrix/Normalized confusion matrix obtained by evaluating the classifier in the corresponding testing set after running the experiments 50 times.}
		\label{tab:confusion_matrix}
	\end{table}
	% ([[53821,  3552,   763],
	%       [ 3116, 46620,  1380],
	%       [    0,   356,  2712]])
	% sensitivity [0.94007156 0.91658189 0.89830508]
	% specificity [0.93570058 0.94052676 0.99143265]
	\begin{table}[!h]
		\centering
		\begin{tabular}{c|ccc}
			\hline
			Metrics & No Finding & Lung Opacity & Covid-19  \\
			\hline
			Sensitivity & $92.6\pm1.2\%$ & $91.2\pm1.0\%$ & $84.4\pm3.2\%$\% \\
			%  \hline
			Specificity & $94.3\pm0.5\%$ & $93.6\pm1.0\%$ & $98.0\pm0.7\%$\%  \\
			\hline
		\end{tabular}
		\caption{Results obtained by the classifier in terms of Sensitivity and Specificity metrics. It is reported the mean and standard desviations statistics.}    
		\label{tab:metrics}
	\end{table}

\begin{figure}
		\centering
		%  No Finding images
		\begin{subfigure}[b]{0.3\textwidth}
			\centering
			\includegraphics[height=0.8\textwidth]{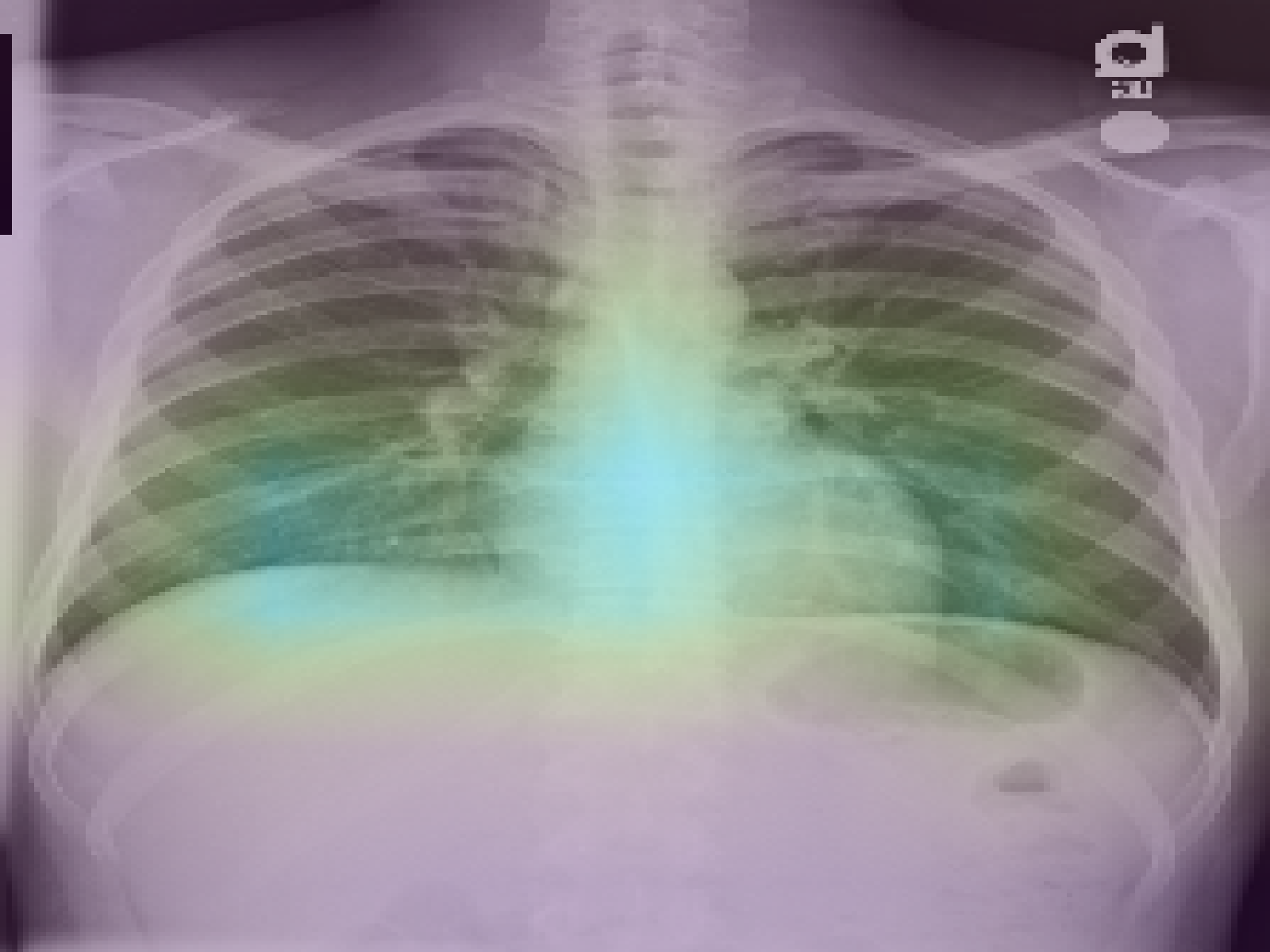}
			\caption{No Finding}
			\label{fig:no_finding1}
		\end{subfigure}
		\hfill
		\begin{subfigure}[b]{0.3\textwidth}
			\centering
			\includegraphics[height=0.8\textwidth]{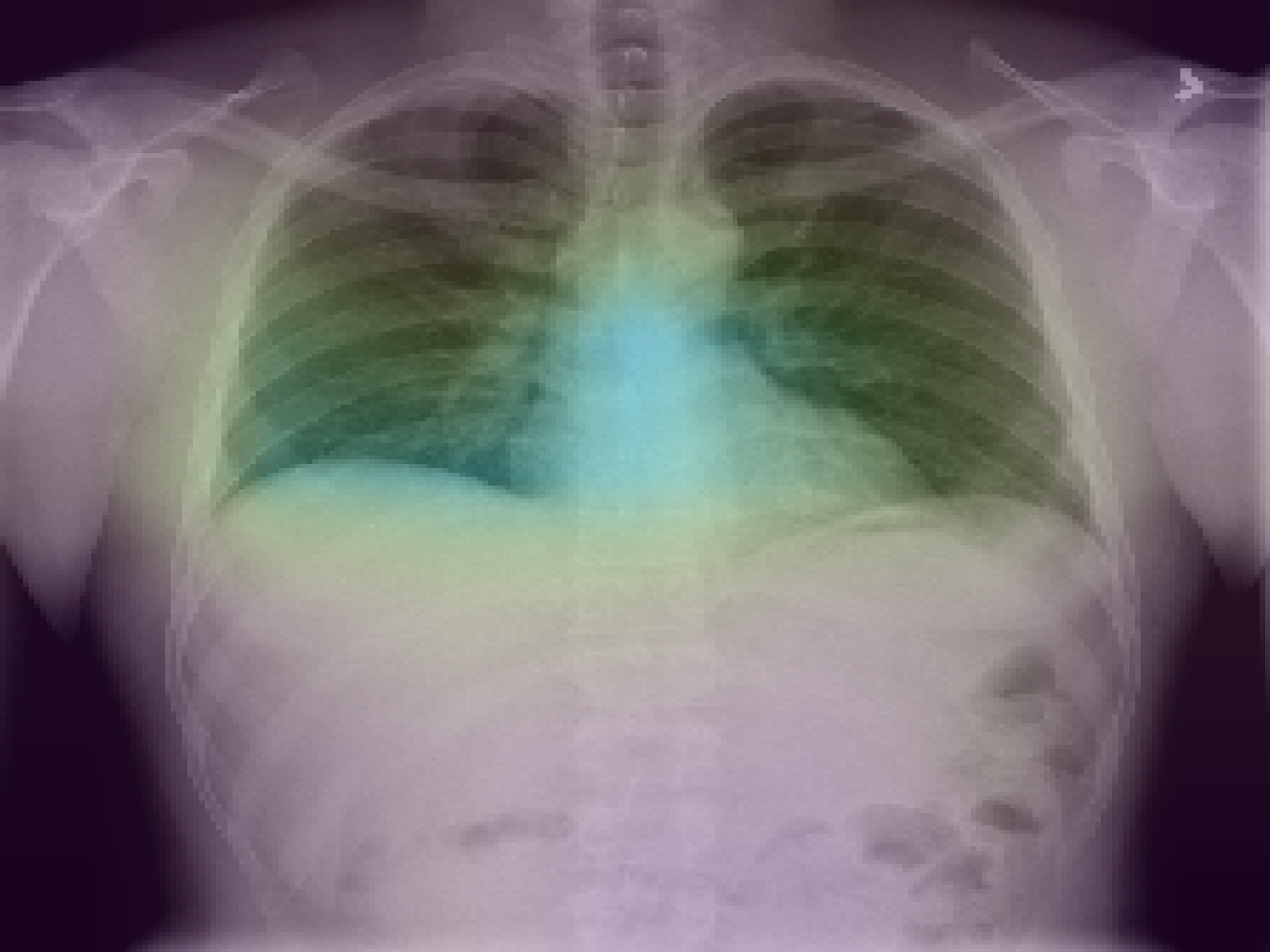}
			\caption{No Finding}
			\label{fig:no_finding2}
		\end{subfigure}
		\hfill
		\begin{subfigure}[b]{0.3\textwidth}
			\centering
			\includegraphics[height=0.8\textwidth]{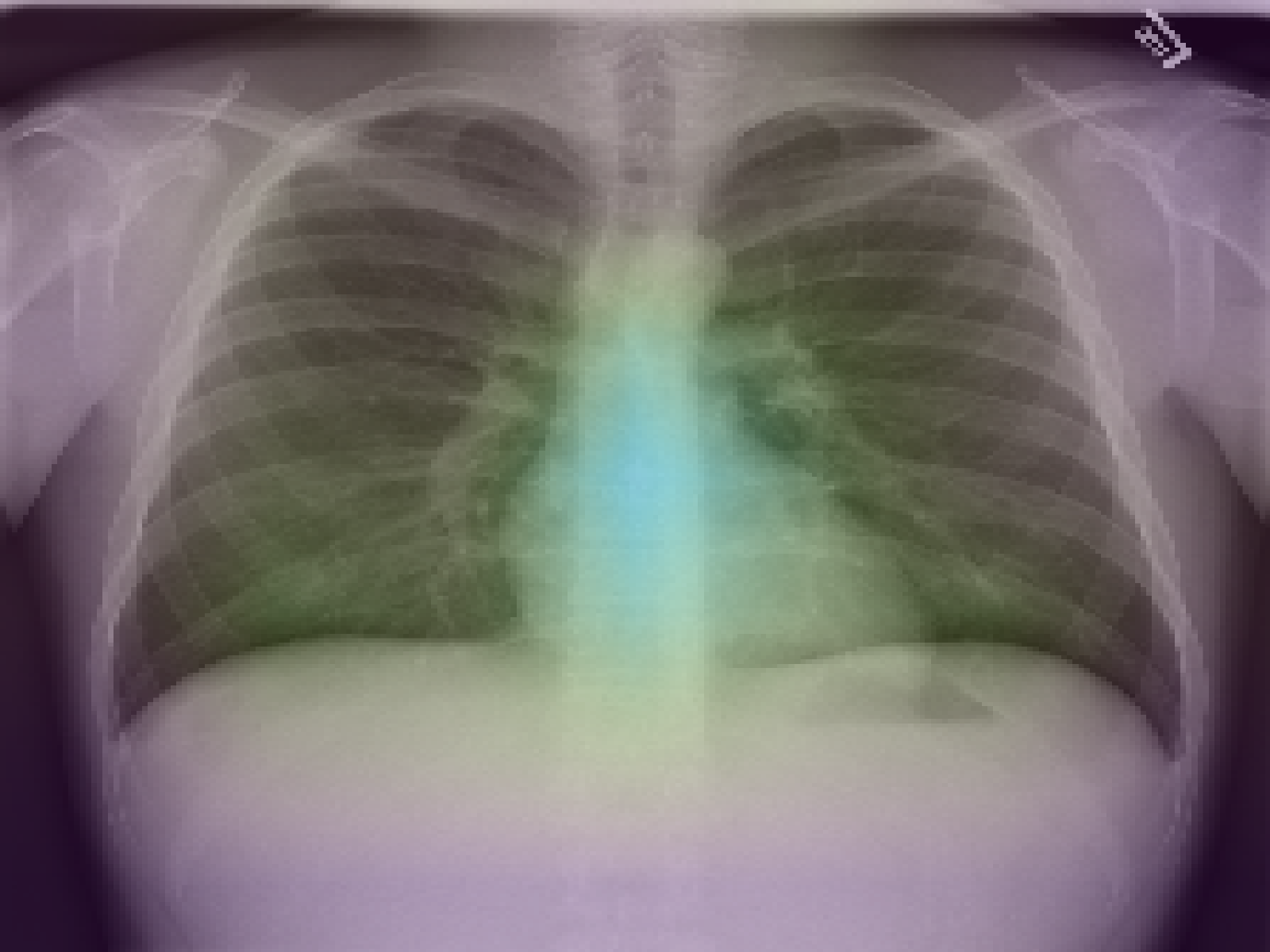}
			\caption{No Finding}
			\label{fig:no_finding3}
		\end{subfigure}
		\vfill
		%  Lung Opacitity
		\begin{subfigure}[b]{0.3\textwidth}
			\centering
			\includegraphics[height=0.8\textwidth]{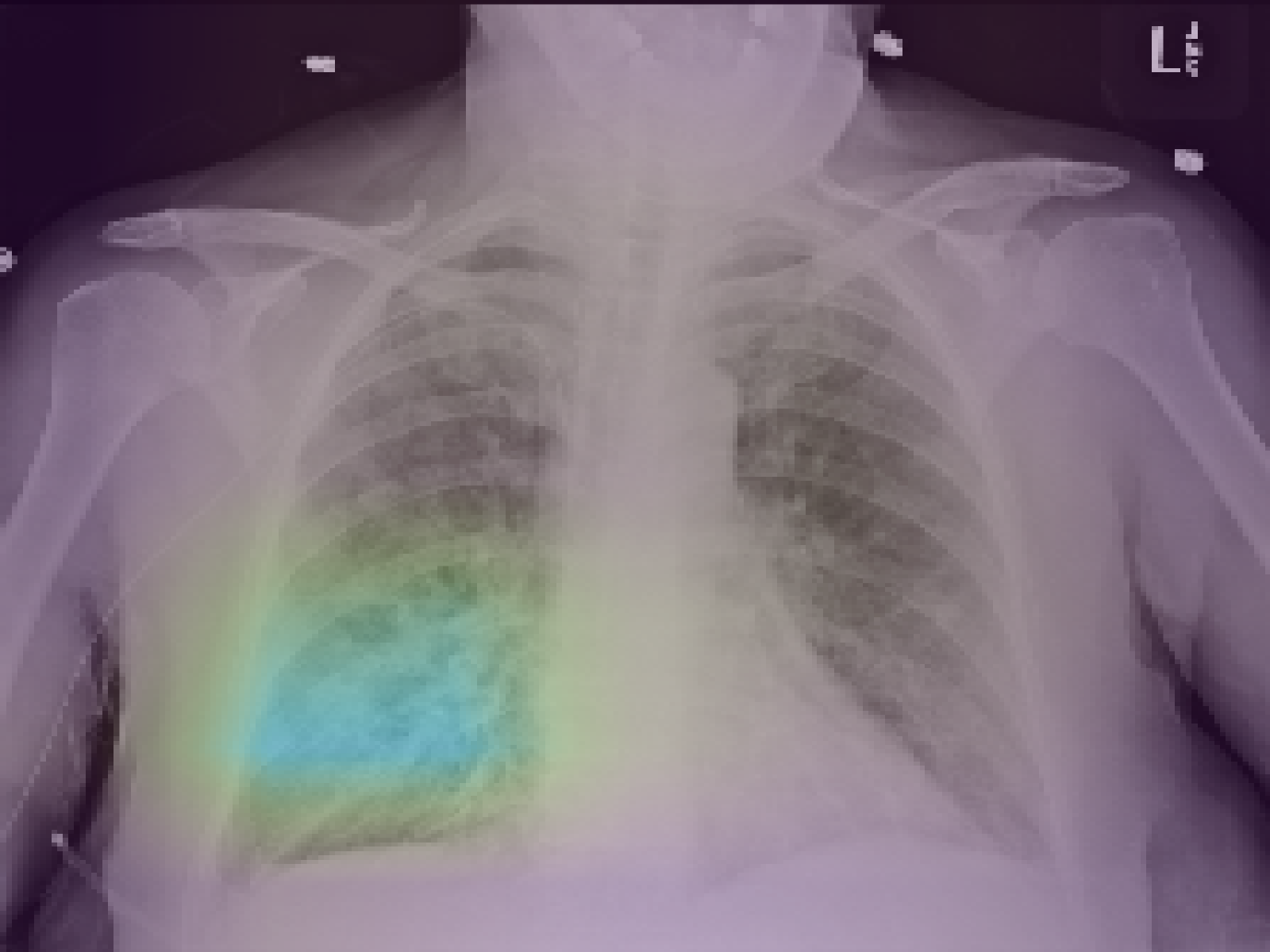}
			\caption{Lung Opacity}
			\label{fig:opacity1}
		\end{subfigure}
		\hfill
		\begin{subfigure}[b]{0.3\textwidth}
			\centering
			\includegraphics[height=0.8\textwidth]{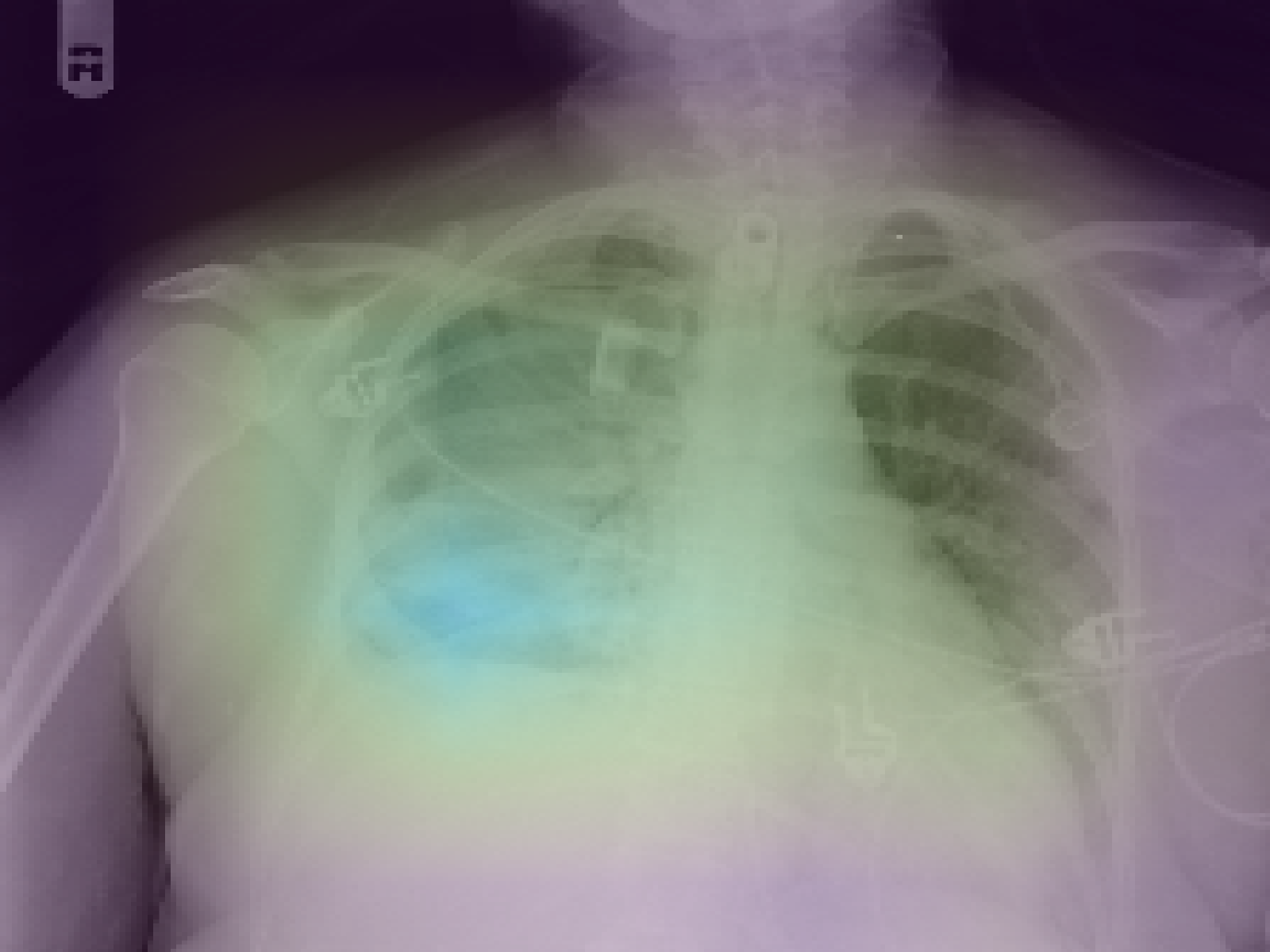}
			\caption{Lung Opacity}
			\label{fig:opacity2}
		\end{subfigure}
		\hfill
		\begin{subfigure}[b]{0.3\textwidth}
			\centering
			\includegraphics[height=0.8\textwidth]{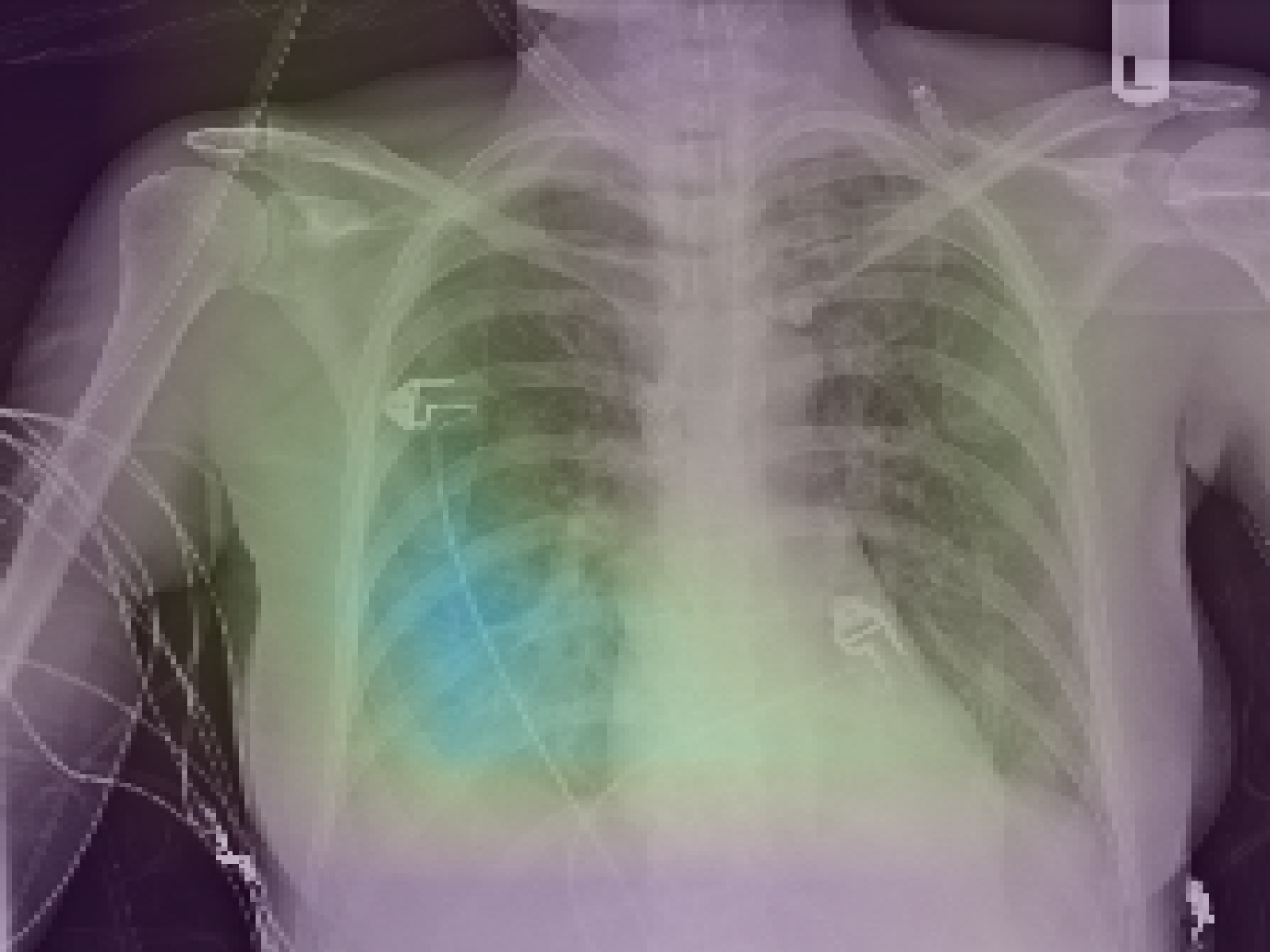}
			\caption{Lung Opacity}
			\label{fig:opacity3}
		\end{subfigure}
		\vfill
		
		\begin{subfigure}[b]{0.3\textwidth}
			\centering
			\includegraphics[height=0.8\textwidth]{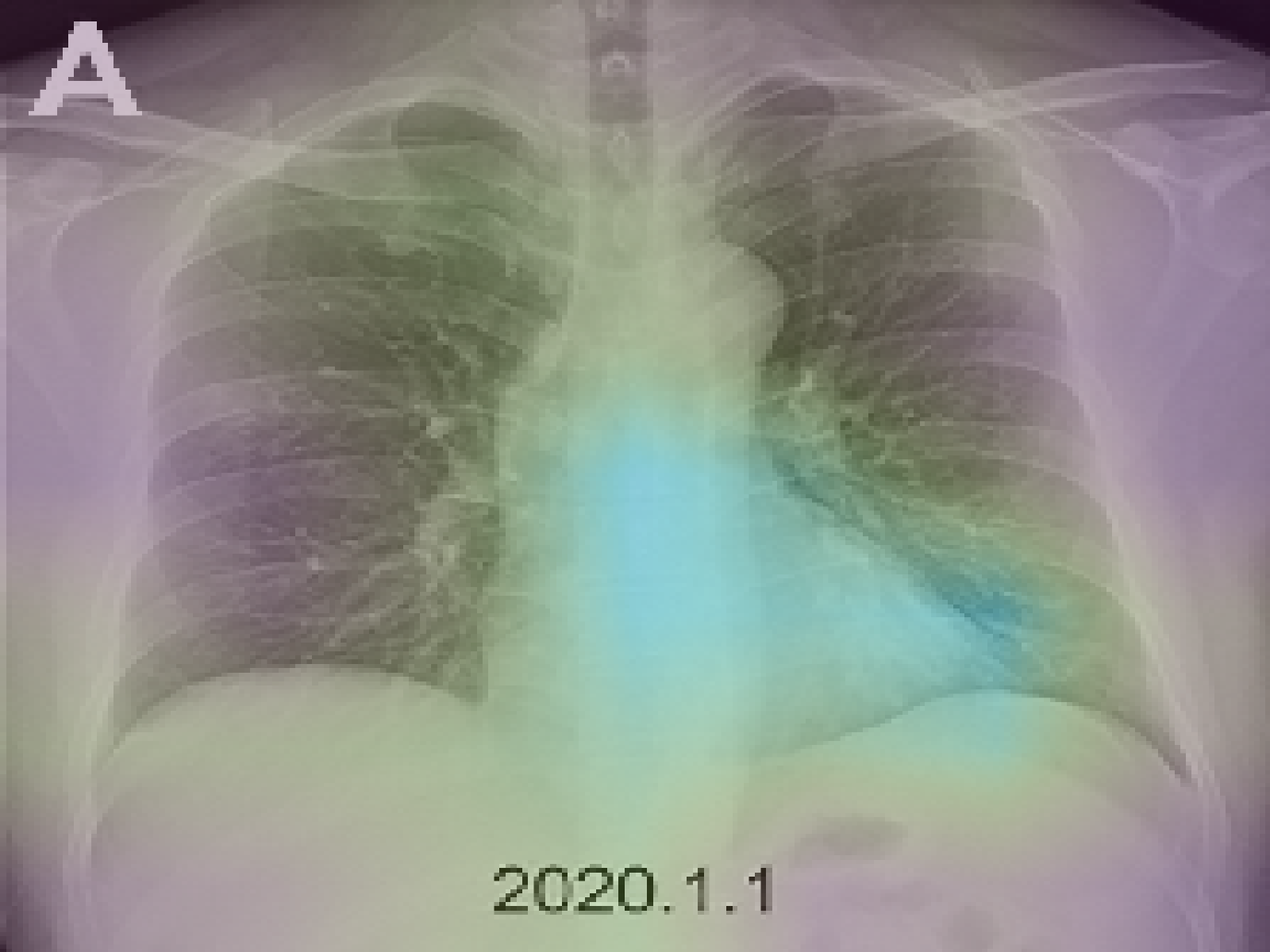}
			\caption{COVID-19}
			\label{fig:covid1}
		\end{subfigure}
		\hfill
		\begin{subfigure}[b]{0.3\textwidth}
			\centering
			\includegraphics[height=0.8\textwidth]{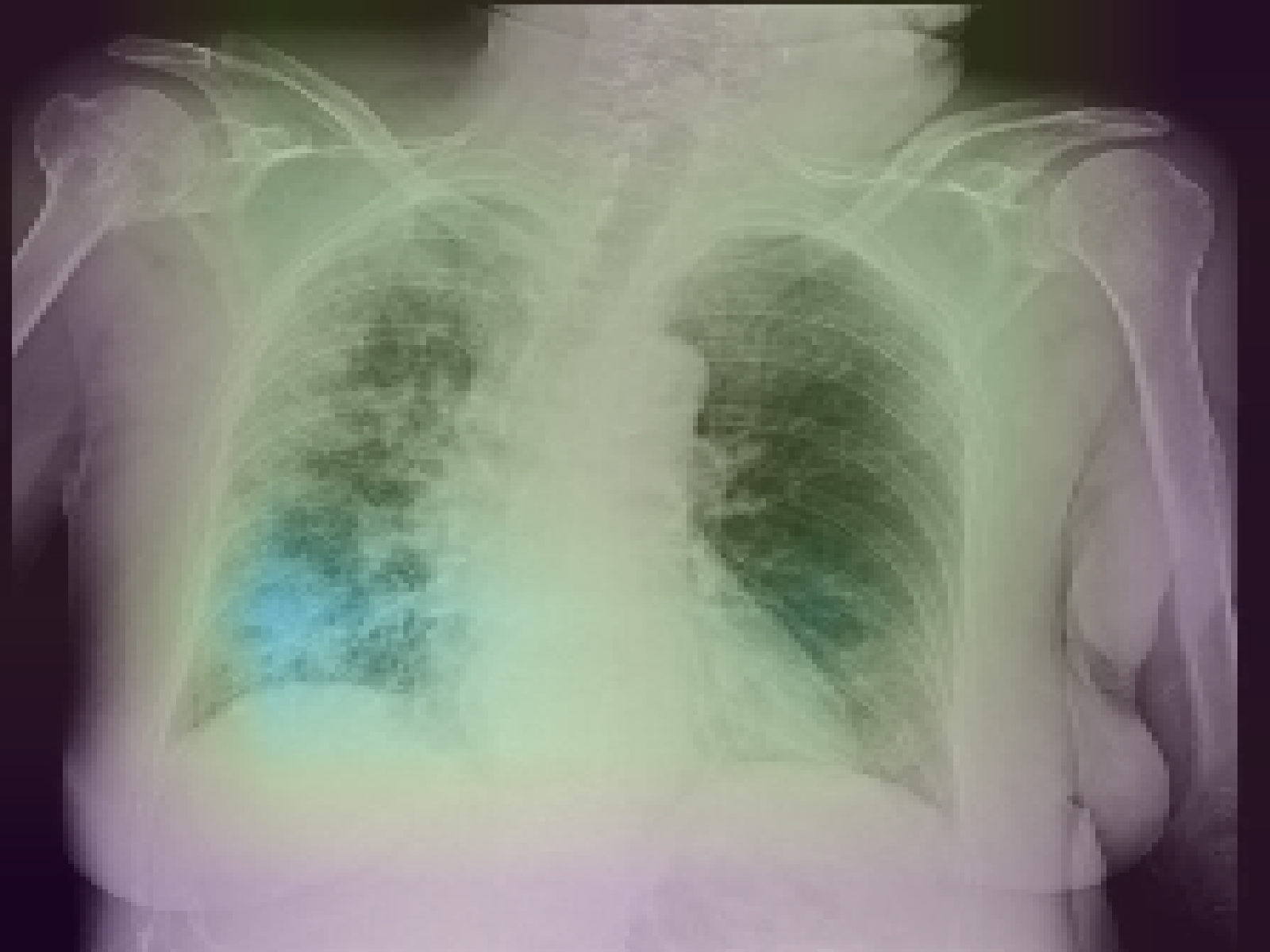}
			\caption{COVID-19}
			\label{fig:covid2}
		\end{subfigure}
		\hfill
		\begin{subfigure}[b]{0.3\textwidth}
			\centering
			\includegraphics[height=0.8\textwidth]{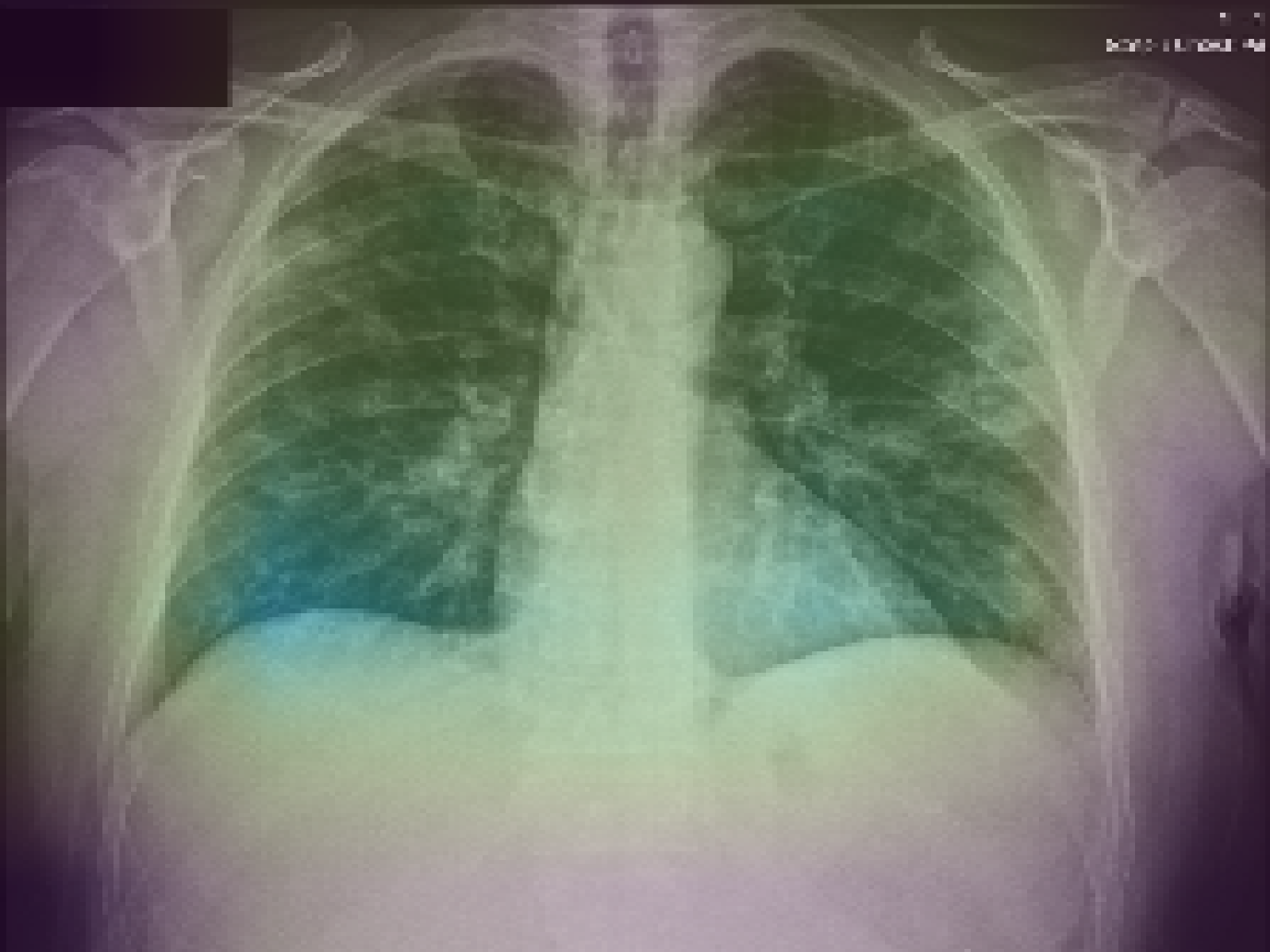}
			\caption{COVID-19}
			\label{fig:covid3}
		\end{subfigure}
		\caption{Examples of images from patients labeled as Non-Finding~(\subref{fig:no_finding1}, \subref{fig:no_finding2}, \subref{fig:no_finding3}), Lung Opacity~(\subref{fig:opacity1}, \subref{fig:opacity2}, \subref{fig:opacity3}), and COVID-19~(\subref{fig:covid1}, \subref{fig:covid2}, \subref{fig:covid3}, respectively)}
		\label{fig:CAMs}
	\end{figure}

	Figure~\ref{fig:CAMs} shows three examples per class of classification activation maps (CAMs) corresponding to images from the testing set. Specifically, it is shown a composition of the images and the associated CAM as heatmaps. By performing this analysis, it can be determined the most relevant regions considered by the network to make the final classification. In fact,  it can be noticed the higher intensities on heatmaps are localized around the lungs as expected, and artifacts as rows or letters present on the images are ignored by the network to make the final prediction. These results indicate that the network is not making the final decision based on the noise characteristics of each dataset. As it is observed in the images of Figure~\ref{fig:covid_examples}, there are artifacts that can easily be identified by the network to discern between images from the assessed classes.
	Regarding images belonging to Lung Opacity, it is remarked that the network was able to identify the regions where these opacities occur. Similar behavior can be observed in the COVID-19 images where the network also highlights the most whited regions. These results are expected considering the high correlation between COVID-19 image and Pneumonia, as COVID-19 causes Pneumonia. For the Non Finding class, the heatmaps are concentrated in the central part of the chest, between the Lungs.
	
\begin{figure}
		\centering
		\includegraphics[width=1.0\textwidth]{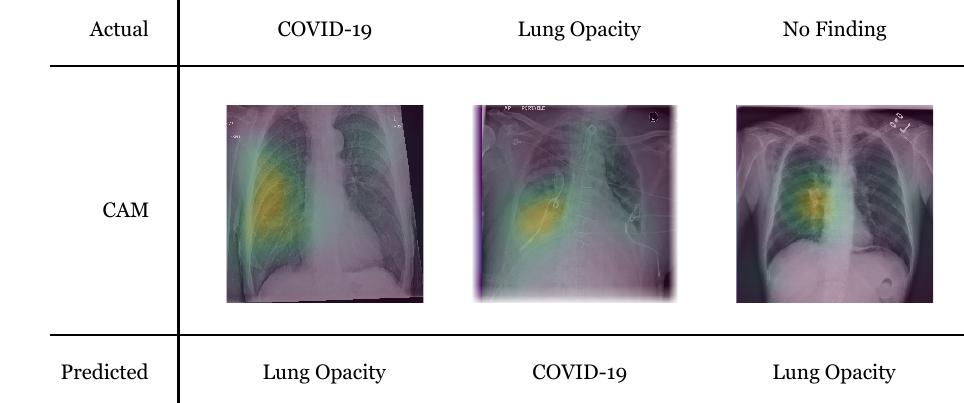}
		\caption{Examples of images missclassified}
		\label{fig:missclassifications}
	\end{figure}

	Finally, Figure~\ref{fig:missclassifications} shows examples of images missclassified. It can be observed that images classified as COVID-19 being of n Lung Opacit, presents similar characteristics and vice-versa.  This situation is expected to be common in patients with advanced degrees of pneumonia. This situation is expected to be common for patients with advanced stages of pneumonia independently the cause of it.
	
	% TO DO: Explain Figure 10

	\subsection{Web Page Interface}
	%O site foi segmentado em três trechos, a parte inicial onde o usuário pode interagir com o site e obter informação relevante focado ao caso em particular, em seguida a segunda seção contendo uma descrição do problema de forma mais ampla para contextualização do problema e, por último mas não menos importante, a região com a apresentação mais detalhada sobre a solução do problema enfrentado e a explicação técnica de como foi colocado em prática. 
	
	The site is structured in three sections, being the first one, where the user can interact with the website to test owned X-Ray images or examples available on the website. The second one describes the motivation and objectives of the project, and the third one presents the proposed solution as well as the technical explanation of its implementation.

	% The site is segmented into three sections, initially a section where the user can be integrated with a site and obtain relevant information focused on the particular case, following the second section that containing a description of the problem in a more comprehensive way to contextualize. Last but not least, the region with the most detailed presentation on the solution of the problem and the technical explanation of how it was placed in practice.
	
	%Na Figura X é exibido o trecho que permanece no início da página para que o usuário, de imediato, possa enviar as imagens e obter uma resposta, sem a necessidade de rolagem até o fim da página. Logo acima da área destinada ao teste do usuário, há uma breve descrição das classes resultantes.
	%Já a Figura X+1 apresenta o histórico de imagens testadas bem como suas avaliações para cada classe (Saudável, Características de Pneumonia e Características de COVID-19) e o diagnóstico.
	
\begin{figure}
		\centering
		\includegraphics[width=0.6\textwidth]{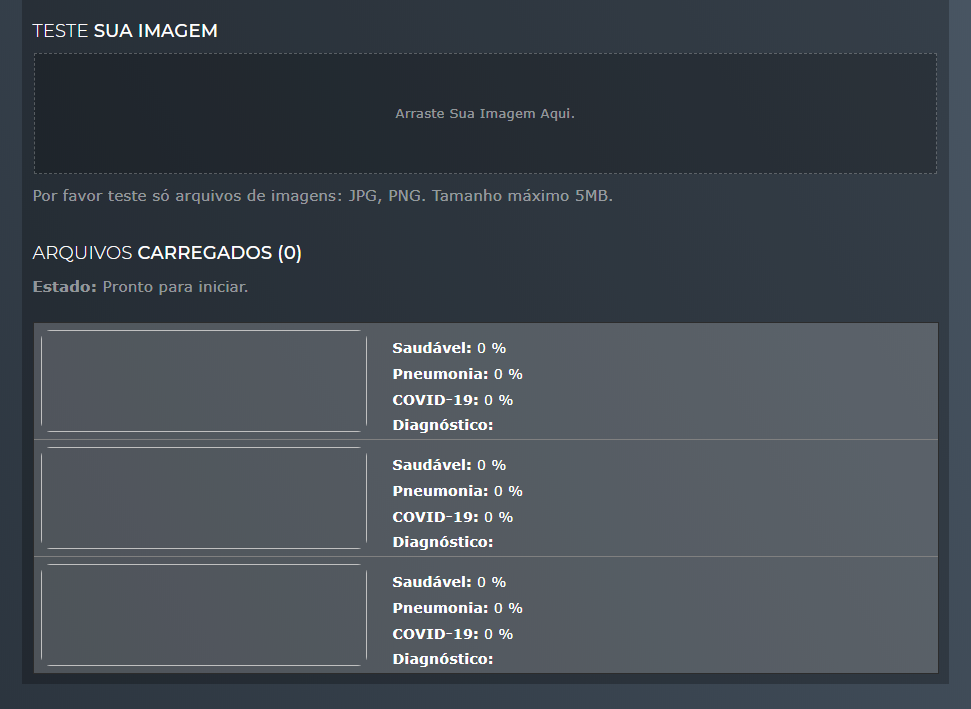}
		\caption{Initial state of the site, where the user can send an X-Ray image to obtain a prediction of COVID-19.}
		%    \caption{Estado inicial da página com o local onde o usuário pode enviar imagens de raio x e obter o prognóstico baseado em cada uma dessas imagens.}
		\label{fig:web1}
	\end{figure}

\begin{figure}
		\centering
		\includegraphics[width=0.6\textwidth]{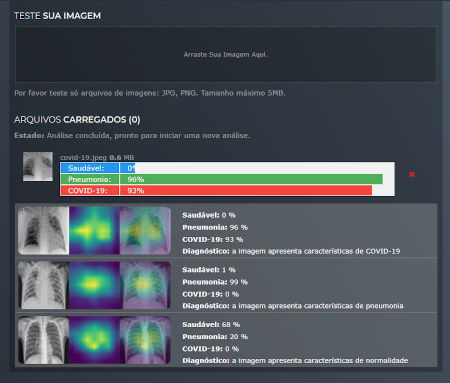}
		\caption{Section of results. History and classification of X-Ray images tested.}
		%Trecho da página onde é exibido o histórico de imagens testadas e seus diagnósticos.}
		\label{fig:web2}
	\end{figure}

	Figure~\ref{fig:web1} displays a snip of the first section of the web site at its initial state, while Figure~\ref{fig:web2} shows examples of testing performed on it. In particular, it is illustrated in diagrams of bars the result obtained for the actual test, and a history of the last experiment carried out. Besides, for each test, it is shown a brief description of the results, indicating if the image evaluated is from a healthy person, with opacity in the Lungs, or with characteristics of COVID-19.
	
	% Figure~\ref{fig:web1} shows the initial section of the site where the user can send the images and get the answer without scrolling to the bottom of the page. Logo above the area intended for user testing, there are a brief description of the resulting classes.
	% Figure~\ref{fig:web2} shows or history of images tested as their evaluations for each class (Healthy, Characteristics of Pneumonia and Characteristics of COVID-19).
	
\begin{figure}
		\centering
		\includegraphics[height=0.6\textwidth]{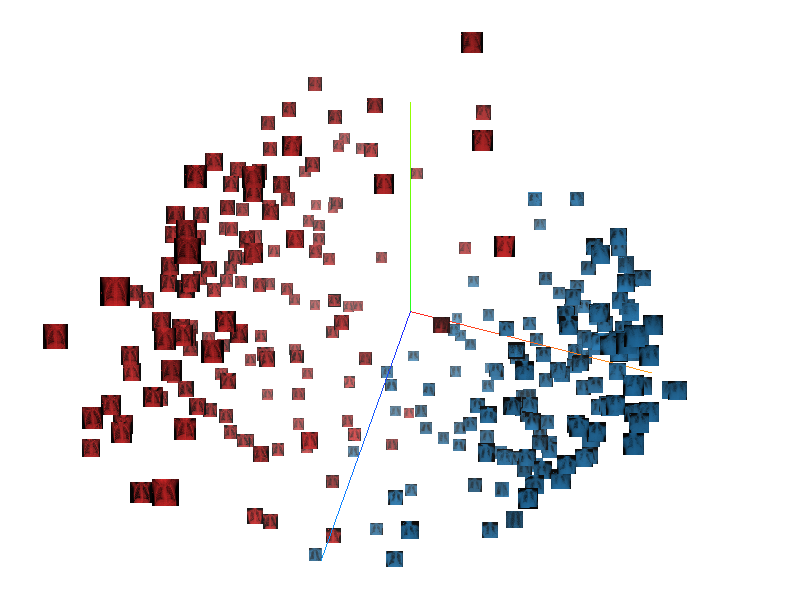}
		\caption{A PCA projection of a dataset to visualized COVID-19 and non-COVID-19 cases.}
		% smilkov_thorat_nicholson
		%Ferramenta interativa disponibilizada pelo Google para ser utilizada pelo usuário para a visualização do agrupamento de imagens.}
		%\footnote{Embedding projector: \url{https://projector.tensorflow.org/}}
		\label{fig:tensorprojetor}
	\end{figure}

	Finally, the web page also shows an Embedding projector~\cite{smilkov2016embedding}\footnote{Embedding projector~\url{https://projector.tensorflow.org/}} to visualize and interpret the features extracted from $294$ X-Ray images from the dataset. For instance, Figure~\ref{fig:tensorprojetor} presents this iterative tool visualizing a basic clustering using the PCA algorithm, where red and blue samples correspond to No Finding and COVID-19 images, respectively 
	
	% Finally, the web page shows and Embedding projector~\cite{smilkov2016embedding}\footnote{Embedding projector~\url{https://projector.tensorflow.org/}} as better visualization and interpretation of embeddings built with $294$ images of our dataset. Figure~\ref{fig:tensorprojetor} presented a iterative tool to visualize basic clustering using PCA algorithm.
	%
	
	\section{Conclusion}
	\label{sec:conclusion}
	% JOSE B.
	% PEDRO
	In this work, a free web service for fast corona virus 2019 (COVID-19) classification using chest X-ray images is presented. The web service is handy and accessible, being possible to use it from desktop computer or mobile devices. It has dedicated GPUs for real-time inference and relies on two deep learning models: a model to differentiate between X-Ray and non-X-Ray images, and a model to detect COVID-19 characteristics in chest X-ray images.
	
	The web service is composed of three main parts: a friendly web interface, an X-Ray image filter, and a COVID-19 classifier. The web interface allows the user to upload a chest X-Ray image for consultation and receive a classification, which can be Healthy, Characteristics of Pneumonia, or COVID-19. Moreover, a heatmap with the Class Activation Maps (CAM) is generated, which indicates the regions that were more relevant in the classification of the evaluated image.
	
	An X-Ray image filter have been implemented to differentiate between valid images (pulmonary frontal X-Ray images), and non-valid images (rotated pulmonary X-Ray images, natural images, other kind of images). A pre-trained MobileNet architecture have been employed for this purpose, achieving an accuracy of 99.3\%.
	
	The COVID-19 Classifier was implemented by optimizing a pre-trained ImageNet DenseNet. Results demonstrated the capability of the model to discern between the assessed classes, recording values above 89\% in terms of sensitivity and specificity metrics. Additionally, a qualitative evaluation, based on the analysis of the generated CAM, indicates that the trained network makes its decisions focuses on the regions of the image where the Lungs are located. This analysis shows that the strategy of augmenting the data and replicating the COVID-19 samples helped to mitigate the problem of overfitting considering the scarcity of COVID-19 samples. 
	
	Future works consider the assessment of the system in hospitals to have feedback from specialists.  We will provide a platform to annotate samples to increase the dataset, and retrain the network to have a more accurate model. 
	
	Besides, we will extend our methodology to make classification in Computerized Tomography images where more COVID-19 images are available.

	% - general conclusion about usage of the whole system: fast, easy to use, covid diagnosis
	% - X-Ray image filter: accuracy, the model works because of ....
	% The network filter to distinguish front X-Ray images
	
	% - COVID-classifier: accuracy, benefits of dense blocks, ...
	% - next steps: test in different hospitals, provide a platform to annotate samples, increase dataset with CT images, make it available as a mobile app
	
	\section*{Acknowledgements}
	
	This work could not have been done without the collaboration of the entire team of the Applied Computational Intelligence Laboratory (ICA) and Cenpes / Petrobras, partners for 20 years in the research and development of artificial intelligence projects for oil and gas sector.
	
	\section*{Disclaimer of liability}
	It is noteworthy that there is no way to guarantee 100~\% effectiveness in any predictive process. For this reason, it is extremely important that any medical diagnosis is made by specialized healthcare professionals. The objective of this project is only to assist decision-making by specialists.
	
	\bibliographystyle{unsrt}  

	\bibliography{main}  %%% Remove comment to use the external .bib file (using bibtex).

	
\end{document}